\documentclass[nofootinbib,twocolumn,eqsecnum]{revtex4}
\usepackage{verbatim}
\usepackage{amssymb}
\usepackage{amsmath}
\usepackage{amsfonts}
\usepackage{bm}
\usepackage{hyperref}
\usepackage{graphicx}
\usepackage{tensor}

\newcommand{\nin}{\noindent}
\newcommand{\nnm}{\nonumber}
\newcommand{\doe}{\partial}
\newcommand{\be}{\begin{equation}}
\newcommand{\ee}{\end{equation}}
\newcommand{\bea}{\begin{eqnarray}}
\newcommand{\eea}{\end{eqnarray}}
\newcommand{\bdm}{\begin{displaymath}}
\newcommand{\edm}{\end{displaymath}}
\newcommand{\bse}{\begin{subequations}}
\newcommand{\ese}{\end{subequations}}

\newcommand{\mr}{\mathrm}
\newcommand{\tr}{\textrm}
\newcommand{\mc}{\mathcal}

\newcommand{\py}{\phantom{yo}}
\newcommand{\qq}{\quad\quad}
\newcommand{\pu}{\phantom{\mu}}
\newcommand{\xg}{\exp}

\begin{document}

\title{Geodesic deviation at higher orders via covariant bitensors}

\author{Justin Vines}
\affiliation{Department of Physics, Cornell University, Ithaca, NY 14853, USA.}

\begin{abstract}

We review a simple but instructive application of the formalism of covariant bitensors, to use a deviation vector field along a fiducial geodesic to describe a neighboring worldline, in an exact and manifestly covariant manner, via the exponential map.  Requiring the neighboring worldline to be a geodesic leads to the usual linear geodesic deviation equation for the deviation vector, plus corrections at higher order in the deviation and relative velocity.  We show how these corrections can be efficiently computed to arbitrary orders via covariant bitensor expansions, deriving a form of the geodesic deviation equation valid to all orders, and producing its explicit expanded form through fourth order.  We also discuss the generalized Jacobi equation, action principles for the higher-order geodesic deviation equations, results useful for describing accelerated neighboring worldlines, and the formal general solution to the geodesic deviation equation through second order.

\end{abstract}

\maketitle

\section{Introduction}\label{sec:intro}

The geodesic deviation equation (GDE), or Jacobi equation, introduced into the canon of basic concepts in general relativity by Synge and Schild \cite{Schild,Synge}, describes the tidal acceleration between neighboring freely-falling observers, and plays an important role in interpreting the curvature of spacetime.
The GDE governs the evolution of a deviation vector field, or Jacobi field, $\xi^\alpha(\tau)$ along a fiducial geodesic worldline $y(\tau)$, which [somehow $\ldots$] specifies a neighboring geodesic worldline $z(\tau)$.  

In elementary textbook treatments \cite{Weinberg,MTW}, one defines the coordinate-basis components of the deviation vector $\xi^\alpha$, at linear order, to be the coordinate differences between points on the worldline $z$ and the fiducial \mbox{geodesic $y$,}
\be\label{xixlinear}
\xi^\alpha(\tau)=z^\alpha(\tau)-y^\alpha(\tau)+O(\xi^2),
\ee
where $\tau$ is an affine parameter along both $y$ and $z$.  It follows, via manipulations of connection coefficients, that the neighboring worldline $z$ will be a geodesic, to linear order, if the vector $\xi^\alpha$ satisfies the well-known linear GDE:
\be\label{gde}
\ddot\xi^\alpha+{R^{\alpha}}_{\beta\gamma\delta}\,u^\beta u^\delta \xi^\gamma=O(\xi,\dot\xi)^2,
\ee
where $u^\alpha=dy^\alpha/d\tau$ is the tangent to the geodesic $y(\tau)$, the Riemann tensor is evaluated along $y(\tau)$, dots denote covariant parameter derivatives $D/d\tau$,   and $O(\xi,\dot\xi)^n$ here denotes terms with $n$ or more factors of $\xi^\alpha$ and/or $\dot\xi^\alpha$.

More sophisticated textbook treatments \cite{Synge,Wald} define the deviation vector in the following covariant manner.  Consider a one-parameter family of geodesics $z(\tau,\epsilon)$, where $\epsilon$ labels the geodesics, and $\tau$ is an affine parameter along each geodesic.  Taking $y(\tau)=z(\tau,0)$ as the fiducial geodesic, one can define deviation vectors $\bar\xi^\alpha$ along $y$ to be the tangents to the curves $\tau=\mr{constant}$:
\be\label{de}
\bar\xi^\alpha(\tau)=\left.\frac{\doe z^\alpha(\tau,\epsilon)}{\doe\epsilon}\right|_{\epsilon=0}.
\ee
It follows, after applying $D^2/d\tau^2$ and using the Ricci identity, that $\ddot{\bar\xi}^\alpha+{R^\alpha}_{\beta\gamma\delta}u^\beta u^\delta \bar\xi^\gamma=0$, like Eq.~(\ref{gde}) but exact.  To connect with the previous paragraph, we can identify, for a given $\epsilon\gtrsim0$, $z(\tau)=z(\tau,\epsilon)$ and $\xi^\alpha=\epsilon\bar\xi^\alpha+O(\epsilon^2)$. 

The generalization of the GDE to higher orders \mbox{in $\xi^\alpha$} seems to have been first considered by Hodgkinson \cite{Hodgkinson}, who, in a nutshell, extended the coordinate-based construction of Eq.~(\ref{xixlinear}) to second and third orders, using manipulations of connection coefficients to derive $O(\xi,\dot\xi)^2$ and $O(\xi,\dot\xi)^3$ corrections to the GDE (\ref{gde}).  A more covariant derivation of the same second-order corrections was subsequently given by Bazanski \cite{Bazanski,Bazanski2}, who, in a nutshell, extended the relation (\ref{de}) by adding to it a term proportional to the second covariant $\epsilon$-derivative of $z(\tau,\epsilon)$.

The basic geometric construction underlying both authors' work, defining the $y$-$\xi^\alpha$-$z$ relationship to all orders, is to reach the point $z$ on the neighboring worldline by following the affinely parametrized geodesic issuing from the point $y$ on the fiducial geodesic with initial tangent vector $\xi^\alpha$ for an affine parameter interval of 1; i.e.,~$z$ is the result of the exponential map of $\xi^\alpha$ at $y$ \cite{Aleksandrov}.  This relationship can be simply and usefully stated in the language of covariant bitensors---developed by Synge \cite{Synge} and DeWitt and Brehme \cite{DeWittBrehme}, succinctly reviewed e.g.~by Poisson et al.~\cite{MPP}, and briefly described below---as
\be\label{link}
\xi^\alpha(\tau)=-\sigma^\alpha\big(y(\tau),z(\tau)\big),
\ee
where $\sigma^\alpha=\nabla^\alpha \sigma$ is the covariant derivative at $y$ of Synge's world function $\sigma(y,z)$.  The link (\ref{link}) between the bitensor formalism and the higher-order geodesic deviations of Hodgkinson and Bazanski seems to have been first noted and employed by Aleksandrov and Piragas \cite{Aleksandrov} in their later (considerably streamlined) rederivation of the second- and third-order GDEs.  

\py

\begin{figure}[h]
\includegraphics[scale=.5]{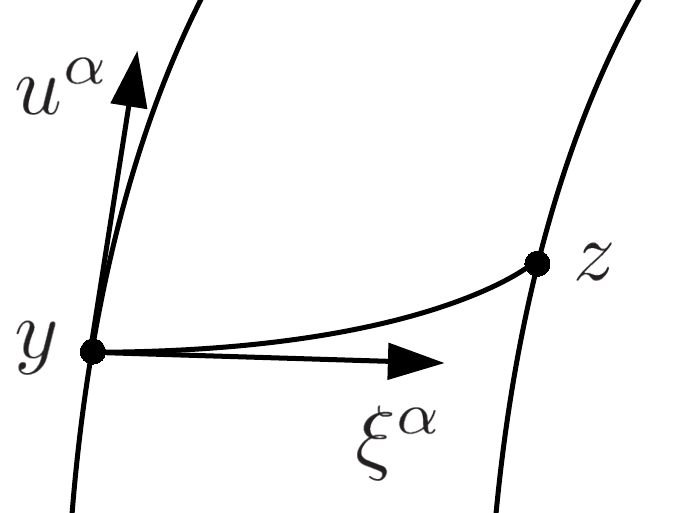}
\label{fig:sigmamu}
\caption{The point $z$ on the neighboring worldline is the exponential map of the deviation vector $\xi^\alpha$ at the point $y$ on the fiducial geodesic.}
\end{figure}

\newpage

One focus of this paper is to expound upon the insights of Ref.~\cite{Aleksandrov}, that one can use bitensor methods to formulate a generalization of the GDE valid to all orders in the deviation, finding its explicit form as a covariant expansion in the deviation vector.  By combining these ideas with the `semi-recursive approach' to bitensor expansions---developed through Refs.~\cite{Avramidi_thesis,Avramidi_book,Decanini,Barry_thesis,Ottewill} for applications to radiation reaction and quantum fields in curved spacetime and quantum gravity---we derive efficient recursion relations which generate the expansion of the GDE, using them to reproduce the results of Refs.~\cite{Hodgkinson,Bazanski,Bazanski2,Aleksandrov} for the second- and third-order GDEs and to produce the fourth-order GDE.  We also rederive results concerning the generalized Jacobi equations of Refs.~\cite{Hodgkinson,Mashhoon75,Mashhoon77,LiNi,Ciufolini,Ciufolini2,Chicone02,Chicone06a,Chicone06b,Perlick} which linearize in the deviation $\xi^\alpha$ but work to all orders in the relative velocity $\dot\xi^\alpha$, and results concerning the description of accelerated neighboring worldlines in terms of covariant deviation vectors, as discussed e.g.~by Refs.~\cite{Synge,Manoff}.

The results for the higher-order GDE are most compactly summarized by the corresponding action functional $S[\xi]=\int d\tau\, \mc L(\xi,\dot\xi)$ for the deviation vector, previously derived through second order [$O(\xi,\dot\xi)^3$ in the \mbox{action}] in Refs.~\cite{Bazanski2,Aleksandrov}, and derived below through fourth order [$O(\xi,\dot\xi)^5$]; the Lagrangian reads
\begin{align} \label{S}
\mc L=&\;
\tfrac{1}{2}\dot\xi^2-\tfrac{1}{2}R_{\xi u\xi u}
\\\nnm&-\tfrac{2}{3}R_{\xi\dot\xi\xi u}-\tfrac{1}{6}R_{\xi u\xi u;\xi}
\\\nnm&
-\tfrac{1}{6}R_{\xi\dot\xi\xi\dot\xi}-\tfrac{1}{4}R_{\xi\dot\xi\xi u;\xi}-\tfrac{1}{24}R_{\xi u\xi u;\xi\xi}-\tfrac{1}{6}R_{\xi u\xi \alpha} {R^{\alpha}}_{\xi\xi u}
\\\nnm&
-\tfrac{1}{12} R_{\xi\dot\xi\xi\dot\xi;\xi}-\tfrac{1}{15}R_{\xi\dot\xi\xi u;\xi\xi}-\tfrac{1}{120}R_{\xi u\xi u;\xi\xi\xi}
\\\nnm&
\qq\quad-\big(\tfrac{2}{15}R_{\xi\dot\xi\xi\alpha}+\tfrac{7}{60}R_{\xi u\xi\alpha;\xi}\big){R^\alpha}_{\xi\xi u}\,+\,O(\xi,\dot\xi)^6,
\end{align}
where, e.g., $R_{\xi\dot\xi\xi u;\xi\xi}=\xi^\alpha \dot\xi^\beta\xi^\gamma u^\delta \xi^\epsilon \xi^\zeta \nabla_\zeta \nabla_\epsilon R_{\alpha\beta\gamma\delta}$.  The Lagrangian (\ref{S}) and the resultant GDE given by Eqs.~(\ref{gde_exact}, \ref{LJIexp4}) below (in a sense\footnote{\label{foot:isoc}The results (\ref{S}), (\ref{solution}), and (\ref{Larb}) apply to the case of `isochronous correspondence,' which requires $\tau$ to be an affine parameter along both the fiducial geodesic $y(\tau)$ and the neighboring worldline $z(\tau)$ defined by Eq.~(\ref{link}); this choice leads to the simplest forms for the GDE and its action.  
\\
$\phantom{abc}$Another useful choice to fix the parametrization of $z(\tau)$ [in the timelike case] is the `normal correspondence,' in which $\xi$ is constrained to be orthogonal to $u$, so that the (parallel-transported tetrad) components of $\xi$, along with $\tau$, correspond Fermi normal coordinates based on the fiducial geodesic (see e.g.~Ref.~\cite{MPP} Sec.~9).  The Lagrangian for the normal correspondence is simply \mbox{$\mc L_\mr{norm.}=\sqrt{1+2\mc L_\mr{isoc.}}-1$}, where $\mc L_\mr{isoc.}$ is the $\mc L$ of Eq.~(\ref{S}); the Lagrangians agree to $O(\xi,\dot\xi)^3$ but differ at higher orders.
\\
$\phantom{abc}$Results for the normal correspondence, as well as for completely generic correspondences/parametrizations are discussed in Appendix \ref{app:gen}, while the main body of the text restricts attention to the isochronous correspondence.}) encapsulate all of the geometric corrections to geodesic deviation considered in Refs.~\cite{Hodgkinson,Bazanski,Bazanski2,Aleksandrov,Mashhoon75,Mashhoon77,LiNi,Ciufolini,Ciufolini2,Chicone02,Chicone06a,Chicone06b,Perlick}.

A second focus of this paper is to review Dixon's construction of the general solution to the linear GDE in terms of fundamental bitensors \cite{Dixon,Dixon1,Dixon3}, and to generalize the construction to find the solution to the second-order GDE.  We show that, given the deviation vector $\xi^\alpha$ and its derivative $\dot\xi^\alpha$ at an initial point $y$ on the fiducial geodesic, the solution for the deviation vector $\xi^{\alpha'}$ at a second point $y'$ on the fiducial geodesic, a finite affine parameter interval $\tau$ along from $y$, can be written as
\begin{align}\label{solution}
\xi^{\alpha'}&={K^{\alpha'}}_\beta\xi^\beta+\tau {H^{\alpha'}}_\beta \dot\xi^\beta
\\\nnm
&\py+\tfrac{1}{2}\left({L^{\alpha'}}_{\beta\gamma}+\tau {H^{\alpha'}}_\alpha{R^\alpha}_{\beta\gamma\delta}u^\delta\right)\xi^\beta\xi^\gamma
\\\nnm&\py+\tau{J^{\alpha'}}_{\beta\gamma}\xi^\beta\dot\xi^\gamma+\tfrac{1}{2} \tau^2 {I^{\alpha'}}_{\beta\gamma} \dot\xi^\beta\dot\xi^\gamma+O(\xi,\dot\xi)^3,
\end{align}
where the `Jacobi propagators' in Dixon's linear solution (the first line) are given in terms of the second derivatives of the world function by
\begin{align}
{K^{\alpha'}}_\beta(y,y')&={H^{\alpha'}}_\gamma(y,y')\;{\sigma^\gamma}_\beta(y,y'),
\nnm\\
{H^{\alpha'}}_\beta(y,y')&=-\Big({\sigma^\beta}_{\alpha'}(y,y')\Big)^{-1},
\end{align}
and the bitensors ${L^{\alpha'}}_{\beta\gamma}$, ${J^{\alpha'}}_{\beta\gamma}$, and ${I^{\alpha'}}_{\beta\gamma}$ in the second-order solution (the last two lines) are given by
\begin{align}
{L^{\alpha'}}_{\beta\gamma}&={K^{\alpha'}}_{\beta;\gamma}+{K^{\alpha'}}_{\beta;\gamma'}{K^{\gamma'}}_{\gamma}\,,
\\\nnm
{J^{\alpha'}}_{\beta\gamma}&={K^{\alpha'}}_{\beta;\gamma'}{H^{\gamma'}}_{\gamma}\,,
\\
{I^{\alpha'}}_{\beta\gamma}&={H^{\alpha'}}_{\beta;\gamma'}{H^{\gamma'}}_{\gamma}\,.\label{LJIintro}
\end{align}

We will see below that the bitensors ${K^{\alpha'}}_\beta$, ${H^{\alpha'}}_\beta$, \mbox{${L^{\alpha'}}_{\beta\gamma}$, ${J^{\alpha'}}_{\beta\gamma}$, and ${I^{\alpha'}}_{\beta\gamma}$} appearing in the solution to the second-order GDE are the same bitensors needed to express the GDE and its action to all orders in the deviation. \{The exact Lagrangian and GDE can be written as
\begin{multline}
\mc L=\tfrac{1}{2}{K^\mu}_\beta K_{\mu\gamma}u^\beta u^\gamma+{K^\mu}_\beta H_{\mu\gamma}u^\beta\dot\xi^\gamma+\tfrac{1}{2}{H^\mu}_{\beta} H_{\mu\gamma}\dot\xi^\beta \dot\xi^\gamma
\\
\Rightarrow\;0={H^\mu}_\alpha \ddot\xi^\alpha+{L^\mu}_{\beta\gamma}u^\beta u^\gamma
+2{J^\mu}_{\beta\gamma}u^\beta \dot\xi^\gamma+{I^\mu}_{\beta\gamma}\dot\xi^\beta \dot\xi^\gamma,\phantom{\Big|}
\label{Larb}
\end{multline}
where the bitensors are functions of $(y,z)$ [or of $(y,\xi)$], and $\mu$ is an index at $z$.\}
In this paper, we define these bitensors in terms of derivatives of the world function $\sigma(y,y')$, and with these definitions they would seem to inherit the limited domain over which $\sigma(y,y')$ and its derivatives are well-defined, requiring the points $y$ and $y'$ to be connected by a unique geodesic segment (to be in each other's normal convex neighborhood (NCN)).  In a companion paper \cite{companion}, we discuss how these and other important bitensors can be alternately defined in terms of the horizontal and vertical covariant derivatives of the exponential map \cite{Dixon,Schattner}, thereby extending the domain of validity of those bitensors and of the solution (\ref{solution}) beyond the NCN, as well as simplifying key expressions and derivations.

\py

We begin in Section \ref{sec:bitensors} with a quick overview of some basic ingredients of the bitensor formalism, including Synge's world function and its derivatives, coincidence limits, the parallel propagator, and covariant expansions near coincidence.  

Section \ref{sec:linear} uses those ingredients to derive, with little effort, the usual linear GDE and its action principle, along with related results which are also applicable to accelerated neighboring worldlines.  

Section \ref{sec:Dixon} derives Dixon's solution to the linear GDE and discusses some properties of the Jacobi propagators.

Section \ref{sec:arb} derives forms of the GDE and its action (and of all the other results given at linear order in Sec.~\ref{sec:linear}) which are valid to all orders in the deviation.  We discuss the expansion of these results first to $O(\xi,\dot\xi)^n$ [as in `Bazanski's equation' at $O(\xi,\dot\xi)^2$] and then to $O(\xi^n)$ but to all orders in $\dot\xi$ [as in the `generalized Jacobi equation' at $O(\xi)$], and we mention various applications of these improved descriptions of geodesic deviation \cite{Mashhoon75,Mashhoon77,LiNi,Ciufolini,Ciufolini2,Chicone02,Chicone06a,Chicone06b,Perlick,Tammelo,Tammelo2,Baskaran,Kerner,vanHolten,Colistete,Colistete2,Koekoek,Koekoek2}.

Appendix \ref{app:gen} extends the analysis of the main text, which considers only the isochronous correspondence, to treat generic correspondences/parametrizations and the normal correspondence.$^{\ref{foot:isoc}}$ 

Appendix \ref{app:ntlo} extends the derivation of the general solution to the GDE to second order in the deviation.

Appendix \ref{app:semi} summarizes results from the semi-recursive/transport-equation approach which efficiently generate high-order expansions of fundamental bitensors.

\section{Bitensors}\label{sec:bitensors}

Bitensors are generalizations of ordinary spacetime tensors which depend on not one but two spacetime points and have a tensor character at each point.  The classic references on the topic are the textbook by Synge \cite{Synge} and the article by DeWitt and Brehme addressing electromagnetic radiation reaction in curved spacetime \cite{DeWittBrehme}.  This section briefly reviews some essential bitensor concepts, borrowing heavily from the thorough review by Poisson et al.~\cite{MPP}.  

To distinguish tensor indices referring to a first point $y$ from those referring to a second point $z$, we use indices $\alpha$, $\beta$, $\gamma$, $\delta$ at $y$ and indices $\mu$, $\nu$, $\rho$ at $z$.  For example, ${K^\mu}_\alpha(y,z)$ is a vector at $z$ and a 1-form at $y$.  (Later, we will add a third point $y'$ with indices $\alpha'$, $\beta'$, $\gamma'$, $\delta'$ and a fourth point $z'$ with indices $\mu'$, $\nu'$, $\rho'$.)

Covariant differentiation can be performed at each of the points $y$ and $z$, denoted by $\nabla_\alpha$ and $\nabla_\mu$ or by semicolons, as in ${K^\nu}_{\beta;\alpha}=\nabla_\alpha {K^\nu}_\beta$ and ${K^\nu}_{\beta;\mu\rho}=\nabla_\rho \nabla_\mu {K^\nu}_\beta$.  For any bitensor field, covariant derivatives at $y$ commute with those at $z$.  For the special case of derivatives of Synge's world function $\sigma(y,z)$, we drop the semicolons, as in ${\sigma^\mu}_{\alpha\beta}=\nabla_\beta\nabla_\alpha\nabla^\mu\sigma$.

\subsection{Synge's world function and its derivatives}\label{sec:Synge}

Consider two points $y$ and $z$ which are connected by a unique geodesic segment $\Gamma$.  The function $\sigma(y,z)$ which gives half the squared geodesic interval along $\Gamma$,
\be\nnm
\sigma(y,z)=\left\{\begin{array}{cccc}-\tfrac{1}{2}(\tr{proper time})^2&\quad&&\tr{timelike}\\0&\quad&&\tr{null}
\\\tfrac{1}{2}(\tr{proper distance})^2&\quad&&\tr{spacelike}\end{array}\right.
\ee
is known as Sygne's world function.  The world function is a biscalar which is symmetric in its arguments.  It is in a sense the fundamental bitensor.

A basic property of $\sigma(y,z)$ is that its first covariant derivatives with respect to $y$ and $z$, $\sigma^\alpha=\nabla^\alpha\sigma$ and \mbox{$\sigma^\mu=\nabla^\mu\sigma$},
yield vectors which are tangent to $\Gamma$ at $y$ and $z$, and their norms are the interval $\sqrt{|2\sigma|}$.  If we parametrize $\Gamma$ as $x=x(\lambda)$, where $\lambda$ is any affine parameter along $\Gamma$, with $x(\lambda_0)=y$ and $x(\lambda_0+\Delta\lambda)=z$, then
\be\label{sigmau}
\sigma^\alpha=-\Delta\lambda\, t^\alpha,\quad \sigma^\mu=\Delta\lambda\,t^\mu,\quad
\sigma =\tfrac{1}{2} (\Delta\lambda)^2\, t^2,\quad
\ee
where $t^\alpha= dx^\alpha/d\lambda(\lambda_0)$ and $t^\mu= dx^\mu/d\lambda(\lambda_0+\Delta\lambda)$ are the tangents to $\Gamma$ at $y$ and $z$ (and $t^2=t^\alpha t_\alpha=t^\mu t_\mu$, because the tangent is parallel-transported along $\Gamma$), as illustrated in Fig.~\ref{fig:sigmamu}.

\begin{figure}[h]
\includegraphics[scale=.48]{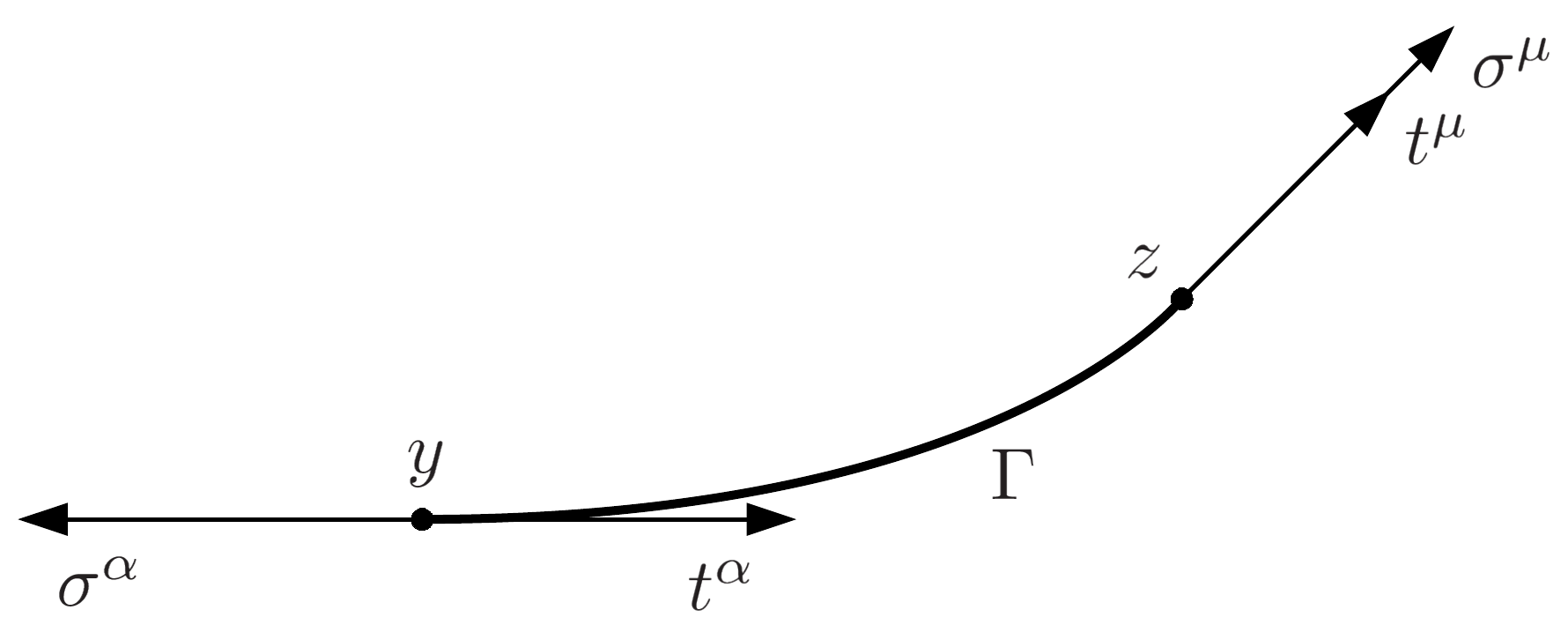}
\caption{The geodesic segment $\Gamma$ linking the points $y$ and $z$.  The tangents, $t^\alpha$ at $y$ and $t^\mu$ at $z$, point in the same direction along $\Gamma$ (here from $y$ to $z$, for $\Delta\lambda>0$).  The world-function derivatives, $\sigma^\alpha$ at $y$ and $\sigma^\mu$ at $z$, both point outward from $\Gamma$.
}\label{fig:sigmamu}
\end{figure}

Rearranging the relations (\ref{sigmau}) yields
\be\label{sigma2}
\sigma^\alpha\sigma_\alpha=2\sigma=\sigma^\mu\sigma_\mu.
\ee
These differential equations, along with the condition $\sigma\to0$ as $z\to y$, completely define the world function.  Differentiating Eqs.~(\ref{sigma2}) yields
\be\label{sigmageod}
\sigma^\beta{\sigma^\alpha}_\beta=\sigma^\alpha,\qquad
\sigma^\nu{\sigma^\mu}_\nu=\sigma^\mu,
\ee
which are geodesic equations for the vector fields $\sigma^\alpha$ (with $z$ fixed) and $\sigma^\mu$ (with $y$ fixed).  Other useful identities can be obtained from further differentiation of Eqs.~(\ref{sigma2}), as in Sec.~4.1 of Ref.~\cite{MPP}.

Note that the point $y$ and the vector $\sigma^\alpha$ at $y$ determine the point $z$: one reaches $z$ by traveling a parameter interval of 1 along the affinely parametrized geodesic issuing from $y$ with initial tangent $-\sigma^\alpha$; i.e., $z$ is the result of the exponential map of $-\sigma^\alpha$ at $y$. The vector $\sigma^\alpha$ (or more appropriately $-\sigma^\alpha$) at $y$ is like a curved-spacetime, covariant version of a displacement vector from $y$ to $z$.

\subsection{Coincidence limits}

The coincidence limit of a bitensor $T_{\alpha\mu\ldots}(y,z)$ is the ordinary tensor at $y$ obtained from the limit $y\to z$, denoted by $[T_{\alpha\mu\ldots}](y)$, where all indices are to be interpreted as indices at $y$.  We assume that all such limits are independent of the path by which $z$ approaches $y$, which will be true for all of the bitensors discussed in this paper as long as the spacetime is sufficiently smooth.

The coincidence limits of the derivatives of the world function in a smooth spacetime follow the pattern$^{\ref{foot_R2}}$ 
\begin{gather}
[\sigma]=0
\nnm\\
\qquad[\sigma_\alpha]=0,\qquad[\sigma_\mu]=0,\qquad
\nnm\\
[{\sigma^\alpha}_\beta]={\delta^\alpha}_\beta,\quad
[{\sigma^\mu}_\nu]={\delta^\mu}_\nu,\quad
[{\sigma^\alpha}_\mu]=-{\delta^\alpha}_\mu,
\nnm\\
[\sigma_{(3)}]=0,
\nnm\\
[\sigma_{(4)}]\sim \mr{Riem},
\nnm\\
[\sigma_{(n>4)}]\sim \nabla^{n-4}\mr{Riem},\label{ddsigma_cl}
\end{gather}
where $\sigma_{(n)}$ stands for any $n$th derivative of $\sigma$ (with derivatives at $y$ and/or $z$) and the last two right-hand sides represent various sums of permutations of the Riemann tensor and its covariant derivatives (evaluated at $y$).  The derivations and results through $n=4$ can be found in Sec.~4 of Ref.~\cite{MPP}, and some higher-order results are given e.g.~in Refs.~\cite{Christensen1,companion}.

\subsection{The parallel propagator}

Given two points $y$ and $z$ linked by a unique geodesic segment $\Gamma$, and given a vector $A^\alpha$ at $y$, consider parallel transporting the vector along $\Gamma$, to obtain the vector $A^\mu$ at $z$.   This is a linear map, 
$$
A^{\mu}\,=\,{g^\mu}_\alpha(y,z)\, A^\alpha,
$$
from vectors at $y$ to vectors at $z$, 
which defines the bitensor ${g^\mu}_\alpha(y,z)$ known as the parallel propagator.

Its inverse ${g^\alpha}_{\mu}$, satisfying ${g^\mu}_\alpha {g^\alpha}_\nu={\delta^\mu}_\nu$ and ${g^\mu}_\alpha {g^\beta}_\mu={\delta^\beta}_\alpha$, is simply ${g^\alpha}_\mu=g_{\mu\nu}g^{\alpha\beta}{g^\nu}_\beta$, so that ordering and raising or lowering of indices do not really matter.  The parallel propagator can also be defined by the differential equations
\be\label{pp}
\sigma^\beta {g^\mu}_{\alpha;\beta}=0=\sigma^{\nu} {g^\mu}_{\alpha;\nu},
\ee
along with the first of the coincidence limits in the pattern$^{\ref{foot_R2}}$
\begin{gather}
[{g^\mu}_{\alpha}]={\delta^\mu}_{\alpha},
\nnm\\
[ {g^\mu}_{\alpha;\beta}]=0,\qquad[ {g^\mu}_{\alpha;\nu}]=0,
\nnm\\
[ {g^\mu}_{\alpha;(2)}]\sim\mr{Riem},
\nnm\\
[ {g^\mu}_{\alpha;(n>2)}]\sim\nabla^{n-2}\mr{Riem}.\label{dpp_cl}
\end{gather}
The coincidence limits through $n=2$ are derived in Sec.~5 of Ref.~\cite{MPP}, and some higher order results can be found in Refs.~\cite{Christensen1,Christensen2,companion}.

\subsection{Expansions near coincidence}\label{sec:exp}

We saw in Sec.~\ref{sec:Synge} how the vector $(-)\sigma^\alpha$ at $y$ is like a covariant version of a displacement vector from $y$ to $z$.  The covariant expansion of bitensors near coincidence (as $z\to y$) expands in powers of the vector $\sigma^\alpha(y,z)$ in analogy to how an ordinary Taylor expansion in flat space expands in powers of a coordinate displacement vector.

A smooth bitensor with indices only at $y$, like $T_{\alpha\beta}(y,z)$, can be expanded as
\be\label{Texp}
T_{\alpha\beta}(y,z)=A_{\alpha\beta}+A_{\alpha\beta\gamma}\sigma^\gamma+\tfrac{1}{2}A_{\alpha\beta\gamma\delta}\sigma^\gamma\sigma^\delta+O(\sigma^\alpha)^3,
\ee
where the $A$'s are ordinary tensors at $y$, determined by coincidence limits and derivatives thereof of $T$ and its derivatives (e.g.~$A_{\alpha\beta}=[T_{\alpha\beta}]$).  For bitensors with indices at $z$, one factor of the parallel propagator is needed for each such index, to turn an expansion of the form (\ref{Texp}) into a bitensor of the proper index structure; for example,
$$
S_{\alpha\mu\nu}(y,z)={g^\beta}_\mu{g^\gamma}_\nu\Big(B_{\alpha\beta\gamma}+B_{\alpha\beta\gamma\delta}\sigma^\delta+O(\sigma^\alpha)^2\Big),
$$
where the $B$'s are ordinary tensors at $y$, determined from $S$ via derivatives and coincidence limits.  Details of the standard (most straightforward) procedures to find the expansion coefficients, which work well at low orders, can be found in Sec.~6 of Ref.~\cite{MPP}.  At higher orders, it becomes increasingly advantageous to employ the semi-recursive (or transport-equation) approach to bitensor expansions, developed by Refs.~\cite{Avramidi_thesis,Avramidi_book,Decanini,Barry_thesis,Ottewill} and utilized below in Appendix \ref{app:semi}.

The bitensors $\sigma^\alpha$ and ${g^\mu}_\alpha$ serve as the basic ingredients of coavriant expansions and cannot be expanded themselves (and similarly for $\sigma$ and $\sigma^\mu$).  The lowest-order (most important) expansions are those of the second derivatives of the world function and the first derivatives of the parallel propagator, which follow the patterns\footnote{\label{foot_R2}At the higher orders in Eqs.~(\ref{ddsigma_cl}), (\ref{dpp_cl}), and (\ref{patt}), terms with multiple factors of the Riemann tensor and its derivatives get mixed into the coincidence limits and expansion coefficients, according to the patterns
$$
\nabla^2\mr{Riem}\sim (\mr{Riem})^2,\quad \nabla^3 \mr{Riem}\sim\mr{Riem}\cdot\nabla\mr{Riem},\quad\ldots
$$
which are seen when one commutes covariant derivatives of the Riemann tensor.  See Eqs.~(\ref{LJIexp4}) for examples of these patterns.}
\begin{align}
\sigma_{(2)}&\;\sim\;\pm\delta+\mr{Riem}\cdot\sigma^2+\ldots+\nabla^{n-2}\mr{Riem}\cdot\sigma^{n}+\ldots
\nnm\\\label{patt}
{g^\mu}_{\alpha;(1)}&\;\sim\; \mr{Riem}\cdot\sigma^1+\ldots+\nabla^{n-1}\mr{Riem}\cdot\sigma^n+\ldots
\end{align}
where $\sigma^n$ represents $n$ factors of the vector $\sigma^\alpha$, and we have omitted the overall factors of the parallel propagator on the right-hand sides.  The patterns for the expansions of higher derivatives of these bitensors follow from a naive application of $\nabla \sigma^1\sim1$.

Some explicit low-order expansions illustrating the patterns (\ref{patt})---those used in the following section to derive the linear GDE and its action---are the world-function second derivatives,
\bea\label{ddsigma}
{\sigma^\alpha}_\beta&=&{\delta^\alpha}_\beta-\tfrac{1}{3}{R^\alpha}_{\gamma\beta\delta}\sigma^\gamma\sigma^\delta+O(\sigma^\alpha)^3,
\\\nnm
{\sigma^\alpha}_\mu&=&-{g^\beta}_\mu \Big({\delta^\alpha}_\beta+\tfrac{1}{6}{R^\alpha}_{\gamma\beta\delta}\sigma^\gamma\sigma^\delta+O(\sigma^\alpha)^3\Big),
\eea
and the parallel-propagator first derivatives,
\bea\label{dg}
{g^\mu}_{\alpha;\beta}&=&\tfrac{1}{2}{g^\mu}_\gamma{R^\gamma}_{\alpha\beta\delta}\sigma^\delta+O(\sigma^\alpha)^2,
\\\nnm
 {g^\mu}_{\alpha;\nu}&=&\tfrac{1}{2}{g^\mu}_\gamma {g^\beta}_\nu {R^\gamma}_{\alpha\beta\delta}\sigma^\delta+O(\sigma^\alpha)^2.
\eea

\section{The linear GDE and its action}\label{sec:linear}

This section applies bitensor methods to derive the usual leading-order GDE and the action principle from which it follows.  Those results easily follow from two more fundamental results, which apply to both geodesic and non-geodesic (accelerated) worldlines $z(\tau)$, namely, the expressions for the tangent and acceleration vectors of the worldline $z(\tau)$ in terms of the deviation vector field $\xi^\alpha(\tau)$ along the fiducial geodesic $y(\tau)$.  The strategy here is to expand in powers of the deviation vector as soon as possible and manipulate expanded expressions.  We will revisit these derivations in Sec.~\ref{sec:arb}, where the strategy will be to find relations valid to all orders in the deviation and then expand the final results.

\begin{figure}[h]
\includegraphics[scale=.5]{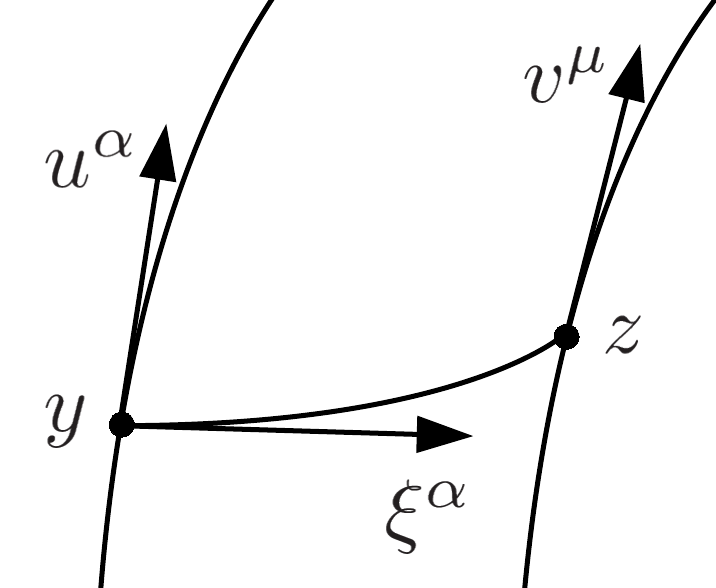}
\caption{The point $z(\tau)$ on the neighboring worldline is the exponential map of the deviation vector $\xi^\alpha(\tau)$ at the point $y(\tau)$ on the fiducial geodesic.
}\label{fig:xi}
\end{figure}

Given a fiducial geodesic $y(\tau)$ and a vector field $\xi^\alpha(\tau)$ along it, we can specify a neighboring worldline $z(\tau)$ in an exact and covariant manner by the relation
\be\label{neighbor}
\xi^\alpha(\tau)=-\sigma^\alpha\big(y(\tau),z(\tau)\big),
\ee
so that $z(\tau)$ is the exponential map of $\xi^\alpha(\tau)$ at $y(\tau)$, as illustrated in Fig.~\ref{fig:xi}.  We denote the tangent to the fiducial geodesic by $u^\alpha= dy^\alpha/d\tau$ and the tangent to the neighboring worldline by $v^\mu=dz^\mu/d\tau$. 

We choose $\tau$ to be an affine parameter along the fiducial geodesic $y(\tau)$.  For a generic vector field $\xi^\alpha(\tau)$, this would imply that the worldline $z(\tau)$ defined by Eq.~(\ref{neighbor}) is non-affinely parametrized.  However, we will restrict attention at first (until Appendix \ref{app:gen}) to the case where $\tau$ is also an affine parameter along $z(\tau)$.  This choice of `isochronous correspondence' [along with Eq.~(\ref{neighbor})] does not uniquely specify a vector field $\xi^\alpha$ given the geodesic $y$ and the worldline $z$, but it specifies a two-parameter family of vector fields related by affine reparametrizations, $\tau\to A\tau+B$, of $z(\tau)$.  We will see below and in Appendix \ref{app:gen} that the `normal correspondence,' in which $\xi$ is constrained to be orthogonal to $u$, is a special case of the isochronous correspondence, for geodesic worldlines $z(\tau)$, to linear order in $\xi$ and $\dot\xi$ (but not at higher orders, and not for accelerated worldlines even at linear order).  Note also that (except where otherwise indicated) our results are equally applicable to timelike, null, and spacelike geodesics.

We use dots to denote covariant $\tau$-derivatives $D/d\tau$; for example, the affinely parameterized geodesic equation for $y(\tau)$ reads $\dot u^\alpha=0$.  We also employ shorthands such as $u^2=u_\alpha u^\alpha$ and $u\cdot\xi=u_\alpha\xi^\alpha$ for squares and contractions of vectors, and ${R^\alpha}_{\beta \xi u}={R^\alpha}_{\beta\gamma\delta}\xi^\gamma u^\delta$, etc., for contractions of vectors with the Riemann tensor (or its derivatives).

Now, consider acting on Eq.~(\ref{neighbor}), $\xi^\alpha=-\sigma^\alpha(y,z)$, with a total covariant $\tau$-derivative:
\bea\label{zdot1}
\dot\xi^\alpha&=&\frac{D}{d\tau}\xi^\alpha=-\frac{D}{d\tau}\sigma^\alpha=-\left(u^\beta\nabla_\beta+v^\mu\nabla_\mu\right)\sigma^\alpha
\nnm\\
&=&-u^\beta{\sigma^\alpha}_{\beta} -v^\mu{\sigma^\alpha}_\mu,
\eea
where we have hidden the $(y,z)$-dependence of the world-function derivatives and the $(\tau)$-dependence of $y$, $z$, $u$, $v$, $\xi$, and $\dot\xi$.  The second derivatives of the world function in Eq.~(\ref{zdot1}) can be covariantly expanded in powers of the deviation vector $\xi^\alpha$, using the coincidence expansion as $z\to y$ in powers of $\sigma^\alpha(y,z)=-\xi^\alpha$, as in Sec.~\ref{sec:exp}.  Inserting Eqs.~(\ref{ddsigma}), with $\sigma^\alpha=-\xi^\alpha$, into Eq.~(\ref{zdot1}) yields
\bea
\dot\xi^\alpha&=&-u^\beta\left({\delta^\alpha}_\beta-\tfrac{1}{3}{R^\alpha}_{\xi\beta\xi}\right)
\\\nnm
&&+\,v^\mu\, {g^\beta}_\mu \left({\delta^\alpha}_\beta+\tfrac{1}{6}{R^\alpha}_{\xi\beta\xi}\right)+O(\xi^3).
\eea
This equation can be solved for $v^\mu$ by working perturbatively in $\xi$ [using the zeroth-order solution $v^\mu={g^\mu}_\alpha (u^\alpha+\dot\xi^\alpha)+O(\xi)$ in the final term], yielding
\bea\label{zdot_exp}
v^\mu&=&{g^\mu}_\alpha\left(u^\alpha+\dot\xi^\alpha-\tfrac{1}{2}{R^\alpha}_{\xi u\xi}\right)
\nnm\\&&+\,O(\xi^3)+\dot\xi\cdot O(\xi^2),
\eea
which expresses the tangent vector $v^\mu$ of the neighboring worldline (which need not be a geodesic) in terms of tensors along the fiducial geodesic ($u$, $\xi$, $\dot\xi$, and the Riemann tensor) and the parallel propagator 
${g^\mu}_\alpha(y,z)$.

We can find an action functional $S[\xi]$ which yields the GDE by expanding the well-known action $S[z]$ which yields geodesic motion for the worldline $z(\tau)$.  The geodesic equation in affine parametrization, $\dot v^\mu=0$, follows from the action $S[z]=\tfrac{1}{2}\int d\tau\, v^2$.  From Eq.~(\ref{zdot_exp}) and $g_{\mu\nu}{g^\mu}_\alpha{g^\nu}_\beta=g_{\alpha\beta}$, the square of the tangent vector is
\bea\label{v2}
v^2&=&u^2+2u\cdot\dot\xi+\dot\xi^2-R_{\xi u\xi u}+O(\xi,\dot\xi)^3.
\eea
Since $u^2$ is a constant and $u\cdot\dot\xi$ is a total derivative (both because $\dot u^\alpha=0$), we can drop those terms from the action, and we obtain
\bea\label{xiSg}
S_\mr{isoc.}[\xi]&=&\int d\tau \Big[\tfrac{1}{2}\dot\xi^2-\tfrac{1}{2}R_{\xi u\xi u} +O(\xi,\dot\xi)^3 \Big].
\eea
Varying this action with respect to $\xi$, using standard techniques, yields the linear GDE:\footnote{The analysis leading to the usual linear GDE (\ref{gde_lin}) requires one to expand in both the `deviation' $\xi$ and the `relative velocity' $\dot\xi$, taking $O(\xi)\sim O(\dot\xi)$, and working to order $O(\xi,\dot\xi)$, meaning dropping all terms with two or more factors of $\xi$ and/or $\dot\xi$.  When $z(\tau)$ is a geodesic, Eq.~(\ref{gde_lin}) implies $\ddot\xi=O(\xi,\dot\xi)$.  However, when $z(\tau)$ is accelerated, generically, $\ddot\xi=O(\xi,\dot\xi)^0$, which means that one must be careful in differentiating expanded expressions [as in Eq.~(\ref{amu})] because in that case $\tfrac{D}{d\tau} [O(\xi,\dot\xi)^n]=O(\xi,\dot\xi)^{n-1}$.}
\be\label{gde_lin}
\ddot\xi^\alpha+{R^\alpha}_{u\xi u}=O(\xi,\dot\xi)^2.
\ee

Note that the quantity $v^2$ of Eq.~(\ref{v2}) is a constant of the motion, because $z(\tau)$ is affinely parametrized.  At $O(\xi,\dot\xi)$, this tells us $D/d\tau(u\cdot\dot\xi)=u\cdot\ddot\xi=O(\xi,\dot\xi)^2$, which also directly follows from the GDE (\ref{gde_lin}).  This implies that $u\cdot\xi=A\tau+B$ is the general solution for the component of $\xi$ parallel to $u$ in the isochronous correspondence at linear order, and the choice $u\cdot\xi=0$ defining the normal correspondence is a consistent special case thereof.

We can also directly calculate the neighboring worldline's normalized acceleration vector $a^\mu$, which in affine parametrization is simply $a_\mr{isoc.}^\mu=\dot v^\mu$ (up to a constant rescaling).  Using Eq.~(\ref{zdot_exp}), and treating $\ddot\xi$ as $O(\xi,\dot\xi)^0$ [as is the case for a generic (accelerated) worldline $z(\tau)$], we have
\bea
a_\mr{isoc.}^\mu&=&\frac{D}{d\tau}\Big[{g^\mu}_\alpha\big(u^\alpha+\dot\xi^\alpha\big)
+O(\xi^2)+O(\xi,\dot\xi)^3\Big]
\nnm\\\label{amu}
&=&{g^\mu}_\alpha\ddot\xi^\alpha+(u^\alpha+\dot\xi^\alpha)\frac{D}{d\tau} {g^\mu}_\alpha
+O(\xi,\dot\xi)^2.
\eea
The $\tau$-derivative of the parallel propagator is given by
$$
\frac{D}{d\tau}{g^\mu}_\alpha=u^\beta{g^\mu}_{\alpha;\beta}+v^\nu {g^\mu}_{\alpha;\nu}
=-{g^\mu}_\beta{R^\beta}_{\alpha u\xi}+O(\xi,\dot\xi)^2,\label{Dg}
$$
where the second equality has used $v^\nu={g^\nu}_\beta u^\beta+O(\xi,\dot\xi)$ from Eq.~(\ref{zdot_exp}) and the expansions (\ref{dg}) of the parallel-propagator derivatives with $\sigma^\alpha=-\xi^\alpha$.  Plugging this into Eq.~(\ref{amu}) yields the expansion
\be\label{amu2}
a_\mr{isoc.}^\mu={g^\mu}_\alpha\Big(\ddot\xi^\alpha+{R^\alpha}_{u\xi u}\Big)
+O(\xi,\dot\xi)^2
\ee
for the acceleration of the neighboring worldline $z(\tau)$.  The geodesic equation $a^\mu=0$ for the worldline then implies the GDE (\ref{gde_lin}) for the deviation vector (because ${g^\mu}_\alpha$ is invertible).

\section{The general solution to the linear GDE}\label{sec:Dixon}

As shown by Dixon \cite{Dixon,Dixon1,Dixon3}, one can write the general solution to the usual linear GDE [the linear GDE (\ref{gde_lin}) for the isochronous (or normal) correspondence] in terms of fundamental bitensors.  While Dixon's original derivation relied on the definition (\ref{de}) of the deviation vector in terms of a one-parameter family of geodesics, this section presents a derivation based on the exponential map definition (\ref{neighbor}).  This construction more easily generalizes to second order in the deviation, which we return to consider in Appendix~\ref{app:ntlo}.

Consider two points, $y$ with indices $\alpha$, $\beta$, etc., and $y'$ with indices $\alpha'$, $\beta'$, etc., linked by a unique geodesic segment which will serve as our fiducial geodesic.  Taking $y$ as a fixed base point, while $y'$ moves along the fiducial geodesic, consider an affine parametrization of the geodesic given by $y'(\tau)$, with $y'(0)=y$, with the tangent $u^{\alpha'}(\tau)=dy^{\alpha'}/d\tau$ at $y'$, and with the tangent $u^\alpha=u^{\alpha'}(0)$ at $y$.  As in Eqs.~(\ref{sigmau}), the tangent $u^\alpha$ at $y$ satisfies
\be\label{rel1}
\tau u^\alpha=-\sigma^\alpha\big(y,y'(\tau)\big).
\ee

\begin{figure}[h]
\includegraphics[scale=.5]{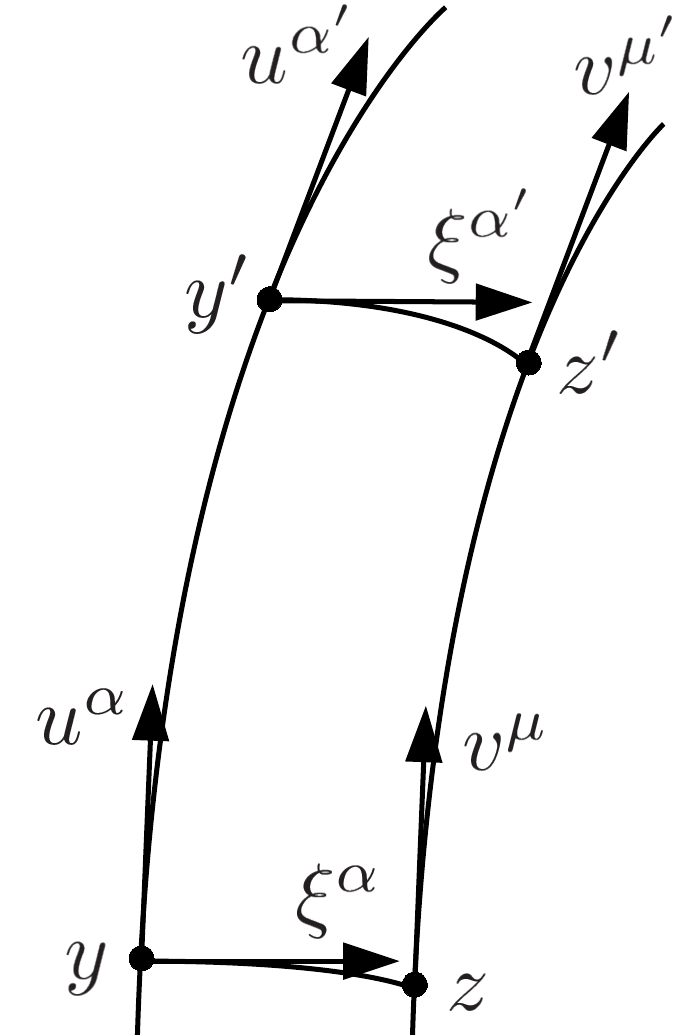}
\caption{The deviation vector $\xi^\alpha$ and its derivative $\dot\xi^\alpha$ at the initial point $y$ on the fiducial geodesic determine the solution for the deviation vector $\xi^{\alpha'}$ at the second point $y'$.
}\label{fig:primes}
\end{figure}

As shown in the previous section, a vector field $\xi^{\alpha'}(\tau)$ along $y'(\tau)$ will satisfy the GDE (\ref{gde_lin}) if the neighboring worldline $z'(\tau)$ defined by
\be\label{zpt}
\xi^{\alpha'}(\tau)=-\sigma^{\alpha'}\big(y'(\tau),z'(\tau)\big)
\ee
is an affinely parametrized geodesic.  We will use indices $\mu'$, $\nu'$, etc.~at the moving points $z'(\tau)$ on the neighboring geodesic, and indices $\mu$, $\nu$, etc.~at the base point \mbox{$z=z'(0)$.}  Because Eq.~(\ref{gde_lin}) is second-order in $\tau$-derivatives, a solution for $\xi^{\alpha'}(\tau)$ is uniquely determined by the initial vector $\xi^\alpha=\xi^{\alpha'}(0)$ at $y$, satisfying
\be\label{rel3}
\xi^\alpha=-\sigma^\alpha(y,z),
\ee
and by the initial derivative $\dot\xi^\alpha=\dot\xi^{\alpha'}(0)$ at $y$.

Denoting the tangent to the neighboring geodesic by $v^{\mu'}(\tau)=dz^{\mu'}/d\tau$, its initial value $v^\mu=v^{\mu'}(0)$ at $z$ must satisfy
\be\label{v_satisfy}
\sigma^\mu(z,z')=-\tau v^\mu= -\tau {g^\mu}_\beta\Big[u^\beta+\dot\xi^\beta+O(\xi,\dot\xi)^2\Big],
\ee
where the first equality follows from Eqs.~(\ref{sigmau}) and the second equality follows from Eq.~(\ref{zdot_exp}).

Consider expanding the function $\sigma^\mu(z,z')$ as $z\to y$ with $z'$ held fixed.  The expansion reads
\begin{align}\label{exp1}
\sigma^\mu(z,z')=&\;{g^\mu}_\beta(y,z)\Big[\sigma^\beta(y,z')-{\sigma^\beta}_\alpha(y,z')\,\sigma^\alpha(y,z)\nnm\\
&+O\big(\sigma^\alpha(y,z)\big)^2\Big],
\end{align}
as can be verified by taking the coincidence limit as $z\to y$ of this equation and its covariant derivative with respect to $z$, using Eqs.~(\ref{ddsigma_cl}), (\ref{pp}), and (\ref{dpp_cl}).  The functions $\sigma^\beta(y,z')$ and ${\sigma^\beta}_\alpha(y,z')$ appearing here can be similarly expanded as $z'\to y'$ with $y$ held fixed, with the results
\begin{align}\label{exp2}
\sigma^\beta(y,z')&=\sigma^\beta(y,y')-{\sigma^\beta}_{\alpha'}(y,y')\,\sigma^{\alpha'}(y',z')\nnm\\&\quad+O\big(\sigma^{\alpha'}(y',z')\big)^2,
\\\nnm
{\sigma^\beta}_\alpha(y,z')&={\sigma^\beta}_\alpha(y,y')+O\big(\sigma^{\alpha'}(y',z')\big).
\end{align}

Inserting the expansions (\ref{exp1}) and (\ref{exp2}) into Eq.~(\ref{v_satisfy}), using the relations (\ref{rel1}), (\ref{zpt}), and (\ref{rel3}), factoring out ${g^\mu}_\beta$, and simplifying yields
$$
-\tau\,\dot\xi^\beta={\sigma^\beta}_{\alpha'}(y,y')\,\xi^{\alpha'}+{\sigma^\beta}_\alpha(y,y')\,\xi^\alpha+O(\xi,\dot\xi)^2.
$$
If the matrix ${\sigma^\beta}_{\alpha'}(y,y')$ is invertible, we can solve this equation to find the desired solution for the deviation vector $\xi^{\alpha'}$ at $y'$ in terms of $\xi^\beta$ and $\dot\xi^\beta$ at $y$:
\be\label{solution_lin}
\xi^{\alpha'}=\,{K^{\alpha'}}_\beta\,\xi^\beta\,+\,\tau\, {H^{\alpha'}}_\beta \,\dot\xi^\beta\,+\,O(\xi,\dot\xi)^2
\ee
where the `Jacobi propagators' are given by
\bea\label{Kdef}
{K^{\alpha'}}_\beta(y,y')&=&{H^{\alpha'}}_\gamma(y,y')\;{\sigma^\gamma}_\beta(y,y'),
\\\label{Hdef}
{H^{\alpha'}}_\beta(y,y')&=&-\Big({\sigma^\beta}_{\alpha'}(y,y')\Big)^{-1},
\eea
with ${}^{-1}$ denoting the matrix inverse.  The matrices $-{\sigma^\beta}_{\alpha'}$ and ${H^{\alpha'}}_\beta$ are those denoted ${D^\beta}_{\alpha'}$ and ${D^{-1\,\alpha'}}_\beta$ by DeWitt and Brehme \cite{DeWittBrehme}, and we have adopted the notation ${K^{\alpha'}}_\beta$ and ${H^{\alpha'}}_\beta$ for the Jacobi propagators from Dixon \cite{Dixon}.

As discussed e.g.~by Refs.~\cite{DeWittBrehme,Harte2}, the matrix $-{\sigma^\beta}_{\alpha'}$ will be invertible as long as the points $y$ and $y'$ are connected by a unique geodesic segment (as long as $y'$ is in the NCN of $y$).  When multiple geodesics connect $y$ to $y'$, generally forming `caustic surfaces,' $-{\sigma^\beta}_{\alpha'}$ becomes divergent (because $\sigma^\beta$ becomes multiple-valued), while its inverse ${H^{\alpha'}}_\beta$ is regular but has one or more zero eigenvalues.  It turns out that the Jacobi propagators ${K^{\alpha'}}_\beta$ and ${H^{\alpha'}}_\beta$, unlike the world-function derivatives ${\sigma^\beta}_{\alpha'}$ and ${\sigma^\beta}_\alpha$, can be extended in a unique and well-defined way beyond the NCN.\footnote{\label{exp_foot}One way to extend the definitions of the Jacobi propagators beyond the NCN is to identify them as the horizontal and vertical covariant derivatives of the exponential map.  This extends their domain of validity, in a nutshell, because the exponential map is well-defined and single-valued over the entire tangent bundle.
Let $\xg(y,\psi)$ denote the exponential map, giving the point (with coordinates $\xg^{\alpha'}$) determined by a point $y$ and a vector $\psi^\alpha$ at $y$, satisfying $\sigma^\alpha\big(y,\xg(y,\psi)\big)=-\psi^\alpha$ in the NCN;  in the context of the above discussion, $\psi^\alpha=\tau u^\alpha$ and $\xg(y,\psi)=y'$; then
\begin{align}
{K^{\alpha'}}_\beta(y,y')&=\nabla_{\beta*} \xg^{\alpha'}(y,\psi)=\left(\frac{\doe}{\doe y^\beta}-\psi^\gamma\Gamma^\delta_{\beta\gamma}\frac{\doe}{\doe\psi^\delta}\right)\xg^{\alpha'},
\nnm\\
{H^{\alpha'}}_\beta(y,y')&=\nabla_{*\beta} \xg^{\alpha'}(y,\psi)=\frac{\doe}{\doe\psi^\beta}\xg^{\alpha'},
\end{align}
where $\nabla_{\beta*}$ and $\nabla_{*\beta}$ denote the horizontal and vertical covariant derivatives of functions on the tangent bundle, introduced by Dixon \cite{Dixon}.  The vertical derivative $\nabla_{*\beta}$ differentiates with respect to the vector $\psi$ at a fixed point $y$, while the horizontal derivative $\nabla_{\beta*}$ differentiates with respect to the point $y$ while parallel-transporting the vector $\psi$ along with it.  We have identified (bitensor) functions of two points $(y,y')$ with functions of a point and a vector at that point $(y,\psi)$, via $(y,y')\to\big(y,\xg(y,\psi)\big)$, which implies
\be
\nabla_{\beta*}=\nabla_\beta+{K^{\beta'}}_\beta \nabla_{\beta'},\quad \nabla_{*\beta}={H^{\beta'}}_\beta\nabla_{\beta'},
\ee
where $\nabla_\beta$ and $\nabla_{\beta'}$ are the usual covariant derivatives with respect to $y$ and $\xg$.  The identity (\ref{commute}) is then seen to be equivalent to the commutativity of the horizontal and vertical covariant derivatives: $$\nabla_{*\gamma}{K^{\alpha'}}_\beta=\nabla_{*\gamma}\nabla_{\beta*} \xg^{\alpha'}=\nabla_{\beta*}\nabla_{*\gamma} \xg^{\alpha'}=\nabla_{\beta*}{H^{\alpha'}}_\gamma.$$  These concepts are motivated and clarified in the companion paper \cite{companion}.}  Considered as functions of the affine parameter $\tau$ along the geodesic $y'(\tau)$, ${K^{\alpha'}}_\beta(y,y'(\tau))$ and ${H^{\alpha'}}_\beta(y,y'(\tau))$ are the unique solutions to the differential equations
\begin{align}
\frac{D^2}{d\tau^2}{K^{\alpha'}}_\beta&={R^{\alpha'}}_{\gamma'\beta'\delta'}u^{\gamma'}u^{\delta'}{K^{\beta'}}_\beta,
\\\nnm
\frac{D^2}{d\tau^2}\left(\tau{H^{\alpha'}}_\beta\right)&={R^{\alpha'}}_{\gamma'\beta'\delta'}u^{\gamma'}u^{\delta'}\left(\tau{H^{\beta'}}_\beta\right),
\end{align}
with the initial conditions
\begin{align*}
\left[{K^{\alpha'}}_\beta\right]=\left[{H^{\alpha'}}_\beta\right]={\delta^{\alpha'}}_\beta,
\\
\left[\frac{D}{d\tau}{K^{\alpha'}}_\beta\right]=\left[\frac{D}{d\tau}{H^{\alpha'}}_\beta\right]=0,
\end{align*}
where the $y'\to y$ coincidence limits (denoted with brackets) correspond to $\tau\to0$.  These equations can be deduced directly from the solution (\ref{solution_lin}) to the linear GDE, or (with some effort) by differentiating the definitions (\ref{Kdef}) and (\ref{Hdef}) and using properties of the world-function derivatives.  Another identity which follows from Eqs.~(\ref{Kdef}) and (\ref{Hdef}) and will be useful in Sec.~\ref{sec:arb} is
\be\label{commute}
{K^{\alpha'}}_{\beta;\gamma'} {H^{\gamma'}}_\gamma = {H^{\alpha'}}_{\gamma;\beta}+{H^{\alpha'}}_{\gamma;\beta'} {K^{\beta'}}_\beta.
\ee

Because the normal correspondence ($u\cdot\xi=0$) is a special case of the isochronous correspondence at this order, the solution (\ref{solution_lin}) also applies to the normal case.  It is instructive to note the identities
\begin{align}
\sigma^\alpha {K^{\alpha'}}_\alpha&=-\sigma^{\alpha'},\qq\qq \sigma_{\alpha'} {K^{\alpha'}}_\alpha=-\sigma_\alpha,
\nnm\\\label{uu}
\Rightarrow\quad u^\alpha {K^{\alpha'}}_\alpha&=u^{\alpha'},\qq\qq\phantom{-} u_{\alpha'} {K^{\alpha'}}_\alpha=u_\alpha,
\end{align}
as well as all four equations with $K\to H$, which follow from Eqs.~(\ref{Kdef}) and (\ref{Hdef}) and derivatives of Eq.~(\ref{sigma2}).  The identities (\ref{uu}) imply that projection orthogonal to the tangent commutes with propagation by $K$ or $H$; i.e.
$$
 {K^{\alpha'}}_\beta {P^{\beta}}_\alpha={P^{\alpha'}}_{\beta'} {K^{\beta'}}_\alpha,\qq {H^{\alpha'}}_\beta {P^{\beta}}_\alpha={P^{\alpha'}}_{\beta'} {H^{\beta'}}_\alpha,
 $$
where ${P^\alpha}_\beta={\delta^\alpha}_\beta+u^\alpha u_\beta$ and ${P^{\alpha'}}_{\beta'}={\delta^{\alpha'}}_{\beta'}+u^{\alpha'} u_{\beta'}$ are the the tensors which project orthogonal to $u^\alpha$ and $u^{\alpha'}$ (in the timelike case).  This means that, if the initial deviation vector $\xi^\alpha$ and the initial derivative $\dot\xi^\alpha$ are orthogonal to $u^\alpha$ at $y$, then the final deviation vector $\xi^{\alpha'}$, the solution to the GDE given by Eq.~(\ref{solution_lin}), will be orthogonal to $u^{\alpha'}$ at $y'$.

\section{The GDE and its action to arbitrary orders}\label{sec:arb}

We now return to consider forms of the GDE and of all the other results of Sec.~\ref{sec:linear} which are valid to all orders in the deviation.  We use all of the same notation and conventions from Sec.~\ref{sec:linear}, and our starting point once again is the exponential map relation,
\be\label{exp_arb}
\xi^\alpha(\tau)=-\sigma^\alpha\big(y(\tau),z(\tau)\big),
\ee
which, given a vector field $\xi^\alpha(\tau)$ along an affinely parametrized fiducial geodesic $y(\tau)$ with tangent $u^\alpha=dy^\alpha/d\tau$,  specifies a neighboring worldline $z(\tau)$ with tangent $v^\mu=dz^\mu/d\tau$, as illustrated in Fig.~\ref{fig:xi} above.  For now, we place no restrictions on the vector field $\xi^\alpha(\tau)$ [except that its norm or extent be small enough to keep $z$ in the NCN of $y$], so that the neighboring worldline $z(\tau)$ is generically non-geodesic and non-affinely parametrized.  

We begin the derivation just as in Sec.~\ref{sec:linear}, by acting on the exponential map relation (\ref{exp_arb}) with a total covariant $\tau$-derivative, yielding the first of Eqs.~(\ref{v_arb}) below.  Solving (exactly) for the neighbor's tangent $v^\mu$, using the definitions (\ref{Kdef}) and (\ref{Hdef}) of the Jacobi propagators ${K^\mu}_\beta(y,z)$ and ${H^\mu}_\beta(y,z)$ in terms of the world-function derivatives ${\sigma^\alpha}_\beta(y,z)$ and ${\sigma^\alpha}_\mu(y,z)$, yields the second equation:
\newpage
\begin{widetext}
\be\label{v_arb}
\dot\xi^\alpha=-{\sigma^\alpha}_\beta u^\beta -{\sigma^\alpha}_\mu v^\mu
\qq\qq\qq\qq\Leftrightarrow\qq\qq\qq\qq
v^\mu={K^\mu}_\beta u^\beta + {H^\mu}_\beta \dot\xi^\beta.
\ee
Acting on each of these relations with another $\tau$-derivative yields
\begin{align*}
\qq\qq
\begin{split}
\ddot\xi^\alpha=&\;-{\sigma^\alpha}_{\beta\gamma}u^\beta u^\gamma - 2{\sigma^\alpha}_{\beta\mu}u^\beta v^\mu 
\\
&\;- {\sigma^\alpha}_{\mu\nu}v^\mu v^\nu -{\sigma^\alpha}_\mu \dot v^\mu
\end{split}
&\qq\qq\Leftrightarrow\qq\qq&
\begin{split}
\dot v^\mu=&\;\big({K^\mu}_{\beta;\gamma}u^\gamma+{K^\mu}_{\beta;\nu}v^\nu\big) u^\beta 
\\
&\;+ \big({H^\mu}_{\beta;\gamma}u^\gamma+{H^\mu}_{\beta;\nu}v^\nu\big)  \dot\xi^\beta+{H^\mu}_\beta \ddot\xi^\beta.
\end{split}
\end{align*}
Taking either of these relations, substituting from Eq.~(\ref{v_arb}) for $v^\mu$ (but not for $\dot v^\mu$), and solving for $\dot v^\mu$, one finds
\be\label{vdot_arb}
\dot v^\mu\,=\,{H^\mu}_\alpha \ddot\xi^\alpha\, + \,{L^\mu}_{\beta\gamma}u^\beta u^\gamma
\,+\,2{J^\mu}_{\beta\gamma}u^\beta \dot\xi^\gamma\,+\,{I^\mu}_{\beta\gamma}\dot\xi^\beta \dot\xi^\gamma\,,
\ee
where the expressions for the bitensors ${L^\mu}_{\beta\gamma}(y,z)$, ${J^\mu}_{\beta\gamma}(y,z)$, and ${I^\mu}_{\beta\gamma}(y,z)$ obtained from the two derivations are
\begin{align}
\begin{split}
{L^\mu}_{\beta\gamma}&={H^\mu}_\alpha\Big({\sigma^\alpha}_{\beta\gamma}+2{\sigma^\alpha}_{\nu(\beta}{K^\nu}_{\gamma)}+{\sigma^\alpha}_{\nu\rho}{K^\nu}_\beta{K^\rho}_\gamma\Big)
\\
{J^\mu}_{\beta\gamma}&={H^\mu}_\alpha {H^\nu}_\gamma \Big({\sigma^\alpha}_{\beta\nu}+{\sigma^\alpha}_{\rho\nu}{K^\rho}_\beta\Big)
\\
{I^\mu}_{\beta\gamma}&={H^\mu}_\alpha {H^\nu}_\beta {H^\rho}_\gamma {\sigma^\alpha}_{\nu\rho}\phantom{\Big(}
\end{split}
&\qq\Leftrightarrow\qq&
\begin{split}
{L^\mu}_{\beta\gamma}&={K^\mu}_{\beta;\gamma}+{K^\mu}_{\beta;\nu}{K^\nu}_{\gamma}\phantom{\Big(}
\\
{J^\mu}_{\beta\gamma}&={K^\mu}_{\beta;\nu}{H^\nu}_{\gamma}\phantom{\Big(}
\\
{I^\mu}_{\beta\gamma}&={H^\mu}_{\beta;\nu}{H^\nu}_{\gamma}\phantom{\Big(}
\end{split},\label{LJI}
\end{align}
\end{widetext}
and we have used the identity (\ref{commute}) in the expression for ${J^\mu}_{\beta\gamma}$  on the right.\footnote{\label{sojp_foot} The bitensors of Eqs.~(\ref{LJI}) can be identified as the second horizontal/vertical derivatives of the exponential map $\xg(y,\xi)=z$:
\begin{align}\label{ddX}
{L^\mu}_{\beta\gamma}&=\nabla_{\gamma*}\nabla_{\beta*} \xg^\mu,\nnm
\\
{J^\mu}_{\beta\gamma}&=\nabla_{*\gamma}\nabla_{\beta*} \xg^\mu,
\\
{I^\mu}_{\beta\gamma}&=\nabla_{*\gamma}\nabla_{*\beta} \xg^\mu=\nabla_{*\beta}\nabla_{*\gamma} \xg^\mu.\nnm
\end{align}
(See Footnote \ref{exp_foot} and Ref.~\cite{companion}.)  
The bitensor ${I^\mu}_{\beta\gamma}$ is symmetric in $\beta$ and $\gamma$, as can be seen from the expression on the left in Eqs.~(\ref{LJI}), while ${J^\mu}_{\beta\gamma}$ and ${L^\mu}_{\beta\gamma}$ are not.  While the derivation of Eq.~(\ref{vdot_arb}) determines only the $\beta\gamma$-symmetric part of ${L^\mu}_{\beta\gamma}$, the non-symmetric definitions given in Eqs.~(\ref{LJI}) have been chosen to match the ${L^\mu}_{\beta\gamma}$ of Eqs.~(\ref{ddX}).}\

The important results (\ref{v_arb}b) and (\ref{vdot_arb}) express the tangent $v^\mu$ to $z(\tau)$ and its covariant $\tau$-derivative $\dot v^\mu$ in terms of the vectors $u^\alpha$, $\dot\xi^\alpha$, and $\ddot\xi^\alpha$ along $y(\tau)$ and the bitensors $K$, $H$, $L$, $J$, and $I$.  These bitensors, functions of $(y,z)$, harbor the dependence on the (undifferentiated) deviation vector $\xi^\alpha=-\sigma^\alpha(y,z)$, and they can be covariantly expanded in powers of $\xi$.  

Before expanding, we can write out the `exact' forms of the GDE and its action.  In the isochronous correspondence, when the worldline $z(\tau)$ defined by Eq.~(\ref{exp_arb}) is affinely parametrized, its acceleration vector is $a^\mu_\mr{isoc.}=\dot v^\mu$ (up to a constant rescaling), and thus, from Eq.~(\ref{vdot_arb}), the geodesic equation for $z(\tau)$, or the GDE for $\xi(\tau)$, reads
\begin{align}\label{gde_exact}
0&=a_\mr{isoc.}^\mu=\dot v^\mu
\\\nnm
&={H^\mu}_\alpha \ddot\xi^\alpha + {L^\mu}_{\beta\gamma}u^\beta u^\gamma
+2{J^\mu}_{\beta\gamma}u^\beta \dot\xi^\gamma+{I^\mu}_{\beta\gamma}\dot\xi^\beta \dot\xi^\gamma.
\end{align}
This equation, along with Eqs.~(\ref{LJI}), is equivalent to Eqs.~(6) of Aleksandrov and Piragas \cite{Aleksandrov}.

The action principle which yields the exact GDE (\ref{gde_exact}) can be found from the action $S[z]=\tfrac{1}{2}\int d\tau\,v^2$ for affinely parametrized geodesic motion; using Eq.~(\ref{v_arb}), we have
\begin{align}\label{S_exact}
&S_\mr{isoc.}[\xi]=\int d\tau\, \mc L_\mr{isoc.}(\xi,\dot\xi)=\tfrac{1}{2}\int d\tau\, v^2,
\\\nnm
&v^2={K^\mu}_\beta K_{\mu\gamma}u^\beta u^\gamma+2{K^\mu}_\beta H_{\mu\gamma}u^\beta\dot\xi^\gamma+{H^\mu}_{\beta} H_{\mu\gamma}\dot\xi^\beta \dot\xi^\gamma.
\end{align}
The Lagrangian is a constant of the motion, because $z(\tau)$ is affinely parametrized.

The exact results (\ref{gde_exact}) and (\ref{S_exact}) can be consistently expanded in two ways: to $O(\xi,\dot\xi)^n$, or to $O(\xi^n)$ but to all orders in $\dot\xi$.  The expansions of the bitensors $K$, $H$, $L$, $J$, and $I$, to the orders needed to derive Bazanski's equation at $O(\xi,\dot\xi)^2$ and the generalized Jacobi equation at $O(\xi)$ (and their actions) are given by
\begin{align}\nnm
{K^\mu}_\beta=&\;{g^\mu}_\alpha\bigg[{\delta^\alpha}_\beta-\tfrac{1}{2}{R^\alpha}_{\xi\beta\xi}-\tfrac{1}{6}{R^\alpha}_{\xi\beta\xi;\xi}
+O(\xi^4)\bigg],
\\\nnm
{H^\mu}_\beta=&\;{g^\mu}_\alpha\bigg[{\delta^\alpha}_\beta-\tfrac{1}{6}{R^\alpha}_{\xi\beta\xi}
+O(\xi^3)\bigg],
\\\nnm
{L^\mu}_{\beta\gamma}=&\;{g^\mu}_\alpha\bigg[{R^\alpha}_{\beta\xi\gamma}+\tfrac{1}{2}\Big({R^\alpha}_{\beta\xi\gamma;\xi}-{R^\alpha}_{\xi\beta\xi;\gamma}\Big)
+O(\xi^3)\bigg],
\\\nnm
{J^\mu}_{\beta\gamma}=&\;{g^\mu}_\alpha\bigg[{R^\alpha}_{\gamma\xi\beta}
+O(\xi^2)\bigg],
\\
{I^\mu}_{\beta\gamma}=&\;{g^\mu}_\alpha\bigg[-\tfrac{2}{3}{R^\alpha}_{(\beta\gamma)\xi}
+O(\xi^2)\bigg].
\label{LJIexp}
\end{align}
Recursion relations which generate these expansions to all orders are presented in Appendix \ref{app:semi}, along with explicit results through the orders necessary to derive the $O(\xi,\dot\xi)^4$ GDE and its action.

The following two subsections discuss some details of the two expansion schemes and mention some physical applications of Bazanski's equation and the generalized Jacobi equation found in the literature.

\subsection{Exapnding to $O(\xi,\dot\xi)^n$; \\Bazanski's equation $(n=2)$}

The $O(\xi,\dot\xi)^n$ expansion scheme treats both the deviation $\xi$ and the relative velocity $\dot\xi$ as small parameters, and assumes furthermore that they are of the same order of smallness.  More precisely, it assumes
$
\dot\xi\sim\xi/\mc R\ll 1,
$
where $\mc R$ measures the spacetime's radius of curvature.  This is the approximation that leads to the usual linear GDE (\ref{gde_lin}) at leading order.

The explicit GDE valid to $O(\xi,\dot\xi)^n$ is found by taking the exact GDE (\ref{gde_exact}), inserting the expansions of ${H^\mu}_\alpha$ and ${L^\mu}_{\beta\gamma}$ to $O(\xi^n)$, of ${J^\mu}_{\beta\gamma}$ to $O(\xi^{n-1})$, and of ${I^\mu}_{\beta\gamma}$ to $O(\xi^{n-2})$, and peeling off an overall factor of ${g^\mu}_\alpha$.

At $O(\xi,\dot\xi)^2$, using the expansions (\ref{LJIexp}), we obtain `Bazanski's equation' \cite{Hodgkinson,Bazanski,Bazanski2,Aleksandrov} in the isochronous correspondence:
\be\label{Beq}
\ddot\xi^\alpha+{R^\alpha}_{u\xi u}+R\indices{_{\xi u}^\alpha_{(\xi;u)}}+2{R^\alpha}_{\dot\xi\xi u}=O(\xi,\dot\xi)^3
\ee
We will see in Appendix \ref{app:gen} that this equation also applies to the normal correspondence if it is projected orthogonal to $u$ on the $\alpha$ index.  For geodesic worldlines $z(\tau)$, through $O(\xi,\dot\xi)^2$, the normal correspondence is a special case of the isochronous correspondence; this is not true at $O(\xi,\dot\xi)^3$ and higher.  The second-order GDE (\ref{Beq}) was first derived by Hodgkinson \cite{Hodgkinson}, and the derivation was later improved by Bazanski \cite{Bazanski,Bazanski2} and Aleksandrov and Piragas \cite{Aleksandrov}.

The GDE through $O(\xi,\dot\xi)^3$, given by Eqs.~(\ref{gde_exact}, \ref{LJIexp4}), was also first derived in Ref.~\cite{Hodgkinson} and later rederived in Ref.~\cite{Aleksandrov}; Eqs.~(\ref{gde_exact}, \ref{LJIexp4}) also present for the first time the $O(\xi,\dot\xi)^4$ GDE.  While all of these results are specialized to the isochronous correspondence, we discuss their generalizations to arbitrary correspondences and their specializations to the normal correspondence in Appendix \ref{app:gen}.

The action functional which yields the $O(\xi,\dot\xi)^n$ GDE can be found by expanding the exact action (\ref{S_exact}) to $O(\xi,\dot\xi)^{n+1}$.  From Eqs.~(\ref{LJIexp}), the Lagrangian for the isochronous Bazanski's equation (\ref{Beq}) is
\be\label{BS}
\mc L_\mr{isoc.}=\tfrac{1}{2}\dot\xi^2-\tfrac{1}{2}R_{\xi u\xi u}
-\tfrac{2}{3}R_{\xi\dot\xi\xi u}-\tfrac{1}{6}R_{\xi u\xi u;\xi}+O(\xi,\dot\xi)^4.
\ee
The Lagrangian through $O(\xi,\dot\xi)^5$ was quoted as Eq.~(\ref{S}) in the introduction.

Bazanski's equation has found several fruitful physical applications.  An early example is the analysis of Tammelo \cite{Tammelo,Tammelo2}, which studied the effects of second-order geodesic deviations in gravitational wave detectors (focusing on Weber-type detectors), emphasizing the fact that second-order effects lead to a net longitudinal force, or pressure, on the detector.  In another application to gravitational wave detectors, Baskaran and Grishchuk \cite{Baskaran} showed that the original data analysis scheme for LIGO, which modeled the detector via the linear GDE, could be off by as much as 10\% because it ignored the second-order ``magnetic component of the gravitational force,'' arising from the fourth term in Eq.~(\ref{Beq}).

Another set of applications of Bazanski's equation is represented by the works of Kerner, van Holten, Colistete, and collaborators \cite{Kerner,vanHolten,Colistete,Colistete2,Koekoek,Koekoek2}.  These works have used higher-order geodesic deviations to generate analytic approximations for mildly eccentric orbits in the Schwarzschild and Kerr spacetimes, terming them ``relativistic epicycles,'' and have applied these methods to derive analytic approximations for the leading-order gravitational waveforms from small bodies in nearly circular orbits around large black holes.

\subsection{Expanding to $O(\xi^n)$;\\ the generalized Jacobi equation $(n=1)$}

The $O(\xi^n)$ expansion scheme assumes small deviations, $\xi/\mc R\ll 1$, but allows arbitrary relative velocities, with $\dot\xi\sim 1$ corresponding to relativistic relative velocities.  At $O(\xi)$, this approximation leads to the `generalized Jacobi equation' (GJE), first derived by Hodgkinson \cite{Hodgkinson} and further developed by Mashhoon and others \cite{Mashhoon75,Mashhoon77,LiNi,Ciufolini,Ciufolini2,Chicone02,Chicone06a,Chicone06b,Perlick}.  

The GJE in the isochronous correspondence can be found by taking the exact GDE (\ref{gde_exact}) and inserting the expansions (\ref{LJIexp}), all to $O(\xi)$, yielding
\be\label{GJE_isoc}
\ddot\xi^\alpha+{R^\alpha}_{u\xi u}+2{R^\alpha}_{\dot\xi\xi u}+\tfrac{2}{3}{R^\alpha}_{\dot\xi\xi\dot\xi}=O(\xi^2).
\ee
This equation can be generalized to higher orders in $\xi$ by using the higher order expansions in Eqs.~(\ref{LJIexp4}) below.  The GJE in the normal correspondence (the form in which it is usually used) differs slightly from Eq.~(\ref{GJE_isoc}) and is given by Eq.~(\ref{GJEnorm}) below.

The action yielding Eq.~(\ref{GJE_isoc}) results from inserting the expansions (\ref{LJIexp}) to $O(\xi^2)$ into the exact action (\ref{S_exact}); the resultant Lagrangian is
\be\label{GJES_isoc}
\mc L_\mr{isoc.}=\tfrac{1}{2}\dot\xi^2-\tfrac{1}{2}R_{\xi u\xi u}
-\tfrac{2}{3}R_{\xi\dot\xi\xi u}-\tfrac{1}{6}R_{\xi\dot\xi\xi\dot\xi}+O(\xi)^3.
\ee
Higher orders can be generated from the expansions (\ref{LJIexp4}), and the Lagrangian for the normal correspondence is given by Eq.~(\ref{GJEnorm}).

The GJE has been applied to a number of astrophysical problems, most notably by Mashhoon and Chicone \cite{Mashhoon75, Mashhoon77,Chicone02,Chicone06a,Chicone06b}.  In Refs.~\cite{Mashhoon75, Mashhoon77}, Mashhoon used the GJE to analyze the tidal dynamics of an extended body in a gravitational field, and to estimate the tidal gravitational radiation emitted by such a body, focusing on orbiting bodies in the Kerr spacetime.  Continuing this line of investigation, Chicone and Mashhoon \cite{Chicone02,Chicone06a,Chicone06b} applied the GJE to collections of test particles in the Kerr spacetime, discussing applications to astrophysical relativistic jets, and to tidal dynamics in plane-wave spacetimes and the de Sitter and G\"odel spacetimes.

Other applications of the GJE include Li and Ni's analysis of the coupling of inertial and gravitational effects for accelerated and rotating observers \cite{LiNi}, Ciufolini and Demianski's prescriptions for measuring spacetime curvature using the GJE \cite{Ciufolini,Ciufolini2}, and Perlick's application of the GJE to null goedesics in plane-wave spacetimes and the Schwarzschild spacetime \cite{Perlick}.

\acknowledgements

The author would like to thank \'Eanna Flanagan, David Nichols, Leo Stein, and Barry Wardell for many helpful discussions, and to acknowledge support from
NSF grants PHY-1068541 and PHY-1404105.

\appendix
\section{\\Generic correspondences/parametrizations \\and the normal correspondence $(u\cdot\xi=0)$}\label{app:gen}

The main text above restricted attention to the isochronous correspondence, in which both the fiducial geodesic $y(\tau)$ and the neighboring worldline $z(\tau)$, related to the deviation vector field by $\xi^\alpha(\tau)=-\sigma^\alpha\big(y(\tau),z(\tau)\big)$, are affinely parametrized.  Relaxing this restriction leads to different forms for the GDE and its action principle.

The key results relating the deviation vector $\xi(\tau)$ and the worldline $z(\tau)$ which are still valid for arbitrary correspondences/parametrizations, and which are valid to all orders in the deviation and relative velocity, are Eqs.~(\ref{v_arb}) and (\ref{vdot_arb}), which express the tangent $v^\mu$ to $z(\tau)$ and its covariant $\tau$-derivative:
\begin{align}\label{B1}
v^\mu&={K^\mu}_\beta u^\beta + {H^\mu}_\beta \dot\xi^\beta,
\\\label{B2}
\dot v^\mu&={H^\mu}_\alpha \ddot\xi^\alpha+{L^\mu}_{\beta\gamma}u^\beta u^\gamma
+2{J^\mu}_{\beta\gamma}u^\beta \dot\xi^\gamma+{I^\mu}_{\beta\gamma}\dot\xi^\beta \dot\xi^\gamma.
\end{align}

The forms of the GDE and its action which apply to arbitrary correspondences can be found quite simply by replacing the geodesic equation (\ref{gde_exact}) for $z(\tau)$ and its action (\ref{S_exact}), which applied when $z(\tau)$ was affinely parametrized, with their reparametrization-invariant versions (assuming $v^\mu$ is not null):
\begin{align}\label{B3}
0=a^\mu&=\frac{1}{\sqrt{|v^2|}}\frac{D}{d\tau}\left(\frac{v^\mu}{\sqrt{|v^2|}}\right),
\\\label{B4}
S[\xi]&=-\int d\tau \sqrt{|v^2|}.
\end{align}
Though we will only consider a few cases explicitly, Eqs.~(\ref{B3}, \ref{B4}) can be expanded to any order in the $O(\xi,\dot\xi)^n$ or $O(\xi^n)$ expansion scheme, via Eqs.~(\ref{B1}, \ref{B2}) and the results of Appendix \ref{app:semi}.

\subsection{The generic case to $O(\xi,\dot\xi)$}\label{gen_lin}

Considering for simplicity the timelike case with $u^2=-1$, expanding the action (\ref{B4}) to $O(\xi,\dot\xi)^2$ yields
\begin{align}\label{B5}
S[\xi]&=\int d\tau \Big[-1+u\cdot\dot\xi
\\\nnm
&\qq\qq+\tfrac{1}{2}\dot\xi^2+\tfrac{1}{2}(u\cdot\dot\xi)^2-\tfrac{1}{2}R_{\xi u\xi u}+O(\xi,\dot\xi)^3  \Big],
\end{align}
which, ignoring the constant and the total derivative, differs from the $S_\mr{isoc.}$ of Eq.~(\ref{xiSg}) by the addition of the $\tfrac{1}{2}(u\cdot\dot\xi)^2$ term.  This term modifies the equation of motion from the usual linear GDE (\ref{gde_lin}) to 
\be\label{B6}
{P^\alpha}_\beta\ddot\xi^\beta+{R^\alpha}_{u\xi u}=O(\xi,\dot\xi)^2.
\ee
where ${P^\alpha}_\beta={\delta^\alpha}_\beta+u^\alpha u_\beta$ is the tensor which projects orthogonal to $u$.  Thus, this generic GDE (\ref{B6}) constrains only the components of $\xi$ orthogonal to $u$, leaving the component along $u$ completely unconstrained.

Note that, in the analysis of Eqs.~(\ref{B5}, \ref{B6}), we have assumed $\ddot\xi=O(\xi,\dot\xi)$, which is true when $z(\tau)$ is a geodesic, as implied by the GDE (\ref{B6}).  When $a^\mu\ne 0$, however, this is not true.  The proper generalization of Eq.~(\ref{amu2}), giving the acceleration vector to $O(\xi,\dot\xi)$, to the generic case, is found by expanding Eq.~(\ref{B5}) while assuming $\ddot\xi=O(\xi,\dot\xi)^0$, yielding
\begin{multline}\label{B7}
a^\mu={g^\mu}_\alpha\Big[(1+2u\cdot\dot\xi){P^\alpha}_\beta\ddot\xi^\beta+{R^\alpha}_{u\xi u}
+(u\cdot\ddot\xi){P^\alpha}_\beta\dot\xi^\beta
\\
+\Big\{\dot\xi\cdot\ddot\xi+(u\cdot\dot\xi)(u\cdot\ddot\xi)\Big\}u^\alpha+O(\xi,\dot\xi)^2\Big].
\end{multline}
Note that setting $a^\mu=0$ still yields the generic GDE (\ref{B6}).

\subsection{The normal correspondence \\and Fermi normal coordinates}

Consider the case where the deviation vector $\xi$ is constrained to be orthogonal to $u$, satisfying $u\cdot\xi=0$, which defines the `normal correspondence'.  In this case, if $\{u^\alpha,e^\alpha_i\}$ with $i=1,2,3$ is an orthonormal tetrad which is parallel-transported along the fiducial geodesic, then 
\be\label{x}
\xi^\alpha= x^i e^\alpha_i,
\ee
and the components $x^i$ (along with $\tau$) correspond to Fermi normal coordinates \cite{MPP} based on the fiducial geodesic.  We will use the notation exemplified by
$
{R^i}_{0j0}=e^i_\alpha u^\beta e^\gamma_j u^\delta {R^\alpha}_{\beta\gamma\delta},
$
for the frame components of the Riemann tensor, where $\{-u_\alpha,e^i_\alpha\}$ is the dual tetrad.  Spatial frame indices will be raised and lowered with the Euclidean 3-metric $\delta_{ij}$.

\subsubsection{The usual linear GDE}

This case can be treated by simply taking the results of Sec.~\ref{gen_lin} for the generic correspondence and replacing $\xi^\alpha$ with Eq.~(\ref{x}).  The action (\ref{B5}) [dropping the first line] and GDE (\ref{B6}) become
\begin{align}
S_\mr{norm.}[x]&=\int d\tau\Big[\tfrac{1}{2}\dot x^i\dot x_i-\tfrac{1}{2}R_{i0j0}x^i x^j+O(x,\dot x)^3\Big]
\\
&\Rightarrow\qq\ddot x^i+{R^i}_{0j0} x^j=O(x,\dot x)^2,\label{B10}
\end{align}
which are straight-forward adaptations of their isochronous counterparts.  As mentioned above, the normal correspondence is a special case of the isochronous correspondence, to $O(\xi,\dot\xi)$, when $z(\tau)$ is a geodesic.  A significant difference with the isochronous case occurs for accelerated worldlines $z(\tau)$; from Eq.~(\ref{B7}), the acceleration vector of $z(\tau)$ in the normal correspondence is given by
\be\label{amu_norm}
a_\mr{norm.}^\mu={g^\mu}_\alpha  \Big[\left(\ddot x^i+{R^i}_{0j0}x^j\right)e^\alpha_i+\dot x^i\ddot x_iu^\alpha+O(x,\dot x)^2\Big],
\ee
which differs from the $a^\mu_\mr{isoc.}$ of Eq.~(\ref{amu2}) by the addition of the final term; this term term does not affect the GDE, as one can verify that $a^\mu_\mr{norm.}=0$ still leads to Eq.~(\ref{B10}).

\subsubsection{Bazanski's equation}

From Eqs.~(\ref{B1}-\ref{B4}, \ref{x}, \ref{LJIexp}), Bazanski's equation and its Lagrangian in the normal correspondence are
\begin{align}\label{Bnorm}
&\mc L_\mr{norm.}=\tfrac{1}{2}\dot x^2-\tfrac{1}{2}R_{i0j0}x^ix^j
\\\nnm
&\qq\qq-\tfrac{2}{3}R_{ijk0}x^i\dot x^j x^k-\tfrac{1}{6}R_{i0j0;k}x^i x^j x^k+O(x,\dot x)^4,
\\\nnm
&\ddot x^i+{R^i}_{0j0}x^j+R\indices{_{j 0}^i_{(k;0)}}x^jx^k+2{R^i}_{jk0}\dot x^j x^k=O(x,\dot x)^3,
\end{align}
which are straight-froward adaptations of Eqs.~(\ref{Beq}, \ref{BS}).  The normal correspondence is still a special case of the isochronous correspondence to this order, but one can verify that this fails to be true at the next order and higher.

\subsubsection{The generalized Jacobi equation}

From Eqs.~(\ref{B1}-\ref{B4}, \ref{x}, \ref{LJIexp}), the GJE and its Lagrangian in the normal correspondence are
\begin{align}
&\mc L=-\sqrt{1-\dot x^2}\bigg[1+\frac{1}{2(1-\dot x^2)}\Big(R_{i0j0}+\tfrac{4}{3}R_{ikj0}\dot x^k
\nnm\\&\qq\qq\qq\qq\qq+R_{ikjl}\dot x^k\dot x^l\Big)x^i x^j+O(x^3)\bigg],
\nnm\\
&\ddot x^i+{R^i}_{0j0}x^j+2{R^i}_{jk0}\dot x^j x^k+\tfrac{2}{3}{R^i}_{jkl}\dot x^j x^k \dot x^l\label{GJEnorm}
\\\nnm
&\qq\qq+\left(2{R}_{0jk0}\dot x^j x^k+\tfrac{2}{3}{R}_{0jkl}\dot x^j x^k \dot x^l\right)\dot x^i=O(x^2).
\end{align}

\section{The general solution to the second-order GDE}\label{app:ntlo}

To derive the general solution to the second-order GDE, Bazanski's equation (\ref{Beq}), in the isochronous correspondence, the setup is precisely the same as in Sec.~\ref{sec:Dixon}, where we derived the solution to the linear GDE; see Figure \ref{fig:primes}.  The equations which define the geometrical relations summarized by Fig.~\ref{fig:primes} \emph{to all orders} are Eqs.~(\ref{rel3}, \ref{rel1}, \ref{zpt}) and the first equality of Eq.~(\ref{v_satisfy}),
\begin{align}\label{A1}
\xi^\alpha&=-\sigma^\alpha(y,z),
\\\nnm
\tau u^\alpha&=-\sigma^\alpha(y,y').
\\\nnm
\xi^{\alpha'}&=-\sigma^{\alpha'}(y',z'),
\\\nnm
\tau v^\mu&=-\sigma^\mu(z,z'),
\end{align}
and the valid-to-all-orders generalization of the second equality of Eq.~(\ref{v_satisfy}), which relates $v^\mu$ to $\dot\xi^\alpha$, is Eq.~(\ref{v_arb}b),
\begin{align}
v^\mu&={K^\mu}_\alpha u^\alpha+{H^\mu}_\alpha \dot\xi^\alpha
\nnm\\
&={g^\mu}_\alpha\left[u^\alpha+\dot\xi^\alpha-\tfrac{1}{2}{R^\alpha}_{\xi u\xi}+O(\xi,\dot\xi)^3\right],
\end{align}
where the second line has expanded to second order, matching Eq.~(\ref{zdot_exp}).

Next, just as in Eqs.~(\ref{exp1}, \ref{exp2}) but now to second order, we expand the function $\sigma^\mu(z,z')$ as $z\to y$ with $z'$ fixed, in powers of $\xi^\alpha=-\sigma^\alpha(y,z)$,
\begin{align}
\sigma^\mu(z,z')&={g^\mu}_\alpha\Big[\sigma^\alpha(y,z')
+{\sigma^\alpha}_\beta(y,z')\,\xi^\beta
\\\nnm
&\qq+\tfrac{1}{2}{\sigma^\alpha}_{\beta\gamma}(y,z')\,\xi^\beta\xi^\gamma
+O(\xi^3)\Big],
\end{align}
and then re-expand the resultant $(y,z')$ bitensors as $z'\to y'$ with $y$ fixed, in powers of $\xi^{\alpha'}=-\sigma^{\alpha'}(y',z')$,
\begin{align}
\sigma^\alpha(y,z')&=\sigma^\alpha(y,y')+{\sigma^\alpha}_{\beta'}(y,y')\,\xi^{\beta'}
\nnm\\\nnm&\qq\qq+\tfrac{1}{2}{\sigma^\alpha}_{\beta'\gamma'}(y,y')\,\xi^{\beta'}\xi^{\gamma'}+O(\xi^3),
\\\nnm
{\sigma^\alpha}_\beta(y,z')&={\sigma^\alpha}_\beta(y,y')+{\sigma^\alpha}_{\beta\gamma'}(y,y')\,\xi^{\gamma'}+O(\xi^2),
\\
{\sigma^\alpha}_{\beta\gamma}(y,z')&={\sigma^\alpha}_{\beta\gamma}(y,y')+O(\xi).\label{An}
\end{align}
Combining Eqs.~(\ref{A1}-\ref{An}) yields the key relation
\begin{align}
&-\tau v^\mu=-\tau {g^\mu}_\alpha\left[u^\alpha+\dot\xi^\alpha-\tfrac{1}{2}{R^\alpha}_{\xi u\xi}+O(\xi,\dot\xi)^3\right]
\\
&=\sigma^\mu(z,z')={g^\mu}_\alpha\Big[-\tau u^\alpha +{\sigma^\alpha}_\beta\xi^\beta+{\sigma^\alpha}_{\beta'}\xi^{\beta'}
\nnm\\\nnm
&+\tfrac{1}{2}{\sigma^\alpha}_{\beta\gamma}\xi^\beta\xi^\gamma+{\sigma^\alpha}_{\beta\gamma'}\xi^\beta\xi^{\gamma'}+\tfrac{1}{2}{\sigma^\alpha}_{\beta'\gamma'}\xi^{\beta'}\xi^{\gamma'}+O(\xi,\dot\xi)^3\Big],
\end{align}
where the arguments of all the second and third derivatives of the world function are $(y,y')$.  This equation can be perturbatively solved for $\xi^{\alpha'}$, using the first-order solution $\phantom{\Big|}\xi^{\alpha'}={K^{\alpha'}}_\beta\xi^\beta+\tau {H^{\alpha'}}_\beta\dot\xi^\beta+O(\xi,\dot\xi)^2$ from Eqs.~(\ref{solution_lin}-\ref{Hdef}) in the last two terms.  The result, the solution to the second order GDE, is
\begin{align}
\xi^{\alpha'}&={K^{\alpha'}}_\beta\xi^\beta+\tau {H^{\alpha'}}_\beta \dot\xi^\beta
\\\nnm
&\py+\tfrac{1}{2}\left({L^{\alpha'}}_{\beta\gamma}+\tau {H^{\alpha'}}_\delta{R^\delta}_{\beta\gamma u}\right)\xi^\beta\xi^\gamma
\\\nnm&\py+\tau{J^{\alpha'}}_{\beta\gamma}\xi^\beta\dot\xi^\gamma+\tfrac{1}{2} \tau^2 {I^{\alpha'}}_{\beta\gamma} \dot\xi^\beta\dot\xi^\gamma+O(\xi,\dot\xi)^3,
\end{align}
where the expressions for the bitensors $L$, $J$, and $I$ resulting from this derivation are as in Eqs.~(\ref{LJI}a) [with the $\mu$-type indices replaced by $\alpha'$-type indices].  Using the alternate expressions from Eqs.~(\ref{LJI}b) yields the form of the solution quoted above as Eqs.~(\ref{solution}-\ref{LJIintro}).

\section{Covariant expansions via the semi-recursive/transport-equation method}\label{app:semi}

We summarize here recursion relations which allow one to efficiently generate high-order covariant expansions of fundamental bitensors.  We import several results from Ottewill and Wardell (OW) \cite{Ottewill,Barry_thesis}, who built on the work of Avramidi \cite{Avramidi_thesis,Avramidi_book} and D\'ecanini and Folacci \cite{Decanini}, deriving the recursion relations from transport equations obeyed by the bitensors.  We have implemented the recursion relations in a Mathematica notebook employing the xTensor package \cite{xTensor}.

Given any bitensor ${T^\mu}_{\beta\ldots}(y,z)$ [using the index conventions of Sec.~\ref{sec:bitensors}], we can define a bitensor ${T^\alpha}_{\beta\ldots}(y,z)$ with all the indices at $z$ parallel-transported to indices at $y$,
\be\label{T1}
{T^\mu}_{\beta\ldots}={g^\mu}_\alpha\ldots {T^\alpha}_{\beta\ldots},
\ee
and ${T^\alpha}_{\beta\ldots}$ can be expanded in powers of the deviation vector $\xi^\alpha\equiv-\sigma^\alpha(y,z)$ as
\be\label{T2}
{T^\alpha}_{\beta\ldots}=\sum_{n=0}^\infty\frac{1}{n!}{T^\alpha}_{\beta\ldots(n)},
\ee
where ${T^\alpha}_{\beta\ldots(n)}$ is some tensor at $y$ contracted with $n$ factors of $\xi^\alpha$:
\be\label{T3}
{T^\alpha}_{\beta\ldots(n)}(y,z)={t^\alpha}_{\beta\ldots\delta_1\ldots\delta_n}(y)\,\xi^{\delta_1}\ldots\xi^{\delta_n}.
\ee
The following subsections present recursion relations for the summands ${T^\alpha}_{\beta\ldots(n)}$ for several key bitensors, with many taken directly from OW and others constructed from results of Ref.~\cite{companion}.  Later recursion relations involve the results of earlier ones.  Note that our summands ${T^\alpha}_{\beta\ldots(n)}$ are $(-1)^n$ times those of OW [c.f.~their Eq.~(2.39) and our Eqs.~(\ref{T1}-\ref{T3}) and $\xi^\alpha=-\sigma^\alpha$], though these signs cancel out in all of the recursion relations, and we have renamed several symbols as noted below.

\subsection{World-function second derivatives and Jacobi propagators}

As in Eq.~(\ref{T1}), we can write the three world-function second derivatives and the two Jacobi propagators (\ref{Kdef}, \ref{Hdef}) in terms of bitensors with indices only at $y$ as
\begin{align}
{\sigma^\mu}_\nu&={g^\mu}_\alpha {g^\beta}_\nu {C^\alpha}_\beta,
\nnm\\
{\sigma^\alpha}_\mu&=-{g^\beta}_\mu {D^\alpha}_\beta,
\nnm\\
{\sigma^\alpha}_\beta&= {E^\alpha}_\beta,
\nnm\\
{H^\mu}_\beta&={g^\mu}_\alpha {H^\alpha}_\beta,
\nnm\\
{K^\mu}_\beta&={g^\mu}_\alpha {K^\alpha}_\beta.
\nnm
\end{align}
We have renamed symbols, OW $\to$ here, as $\xi\to C$, $\eta\to-D$, $\lambda\to E$, $\gamma\to-H$, and $K$ is not discussed by OW.

Defining the quantities
\be
{\kappa^\alpha}_{\beta(n)}={R^\alpha}_{\delta_1\beta\delta_2;\delta_3\ldots\delta_n}\xi^{\delta_1}\ldots\xi^{\delta_n}
\ee
($\mc K\to\kappa$), the recursion relations for $H$, $D$, and $C$, which are OW's Eqs.~(4.7, 4.9, 4.10), and which follow from the transport equation OW (3.15) and the relations OW (4.8, 3.14), are given by
\begin{align}
{H^\alpha}_{\beta(0)}&={D^\alpha}_{\beta(0)}={C^\alpha}_{\beta(0)}={\delta^\alpha}_\beta,
\nnm\\
{H^\alpha}_{\beta(1)}&={D^\alpha}_{\beta(1)}={C^\alpha}_{\beta(1)}=0,
\nnm\\
{H^\alpha}_{\beta(n)}&=-\frac{n-1}{n+1}\sum_{k=0}^{n-2}{n-2 \choose k}{\kappa^\alpha}_{\gamma(n-k)}{H^\gamma}_{\beta(k)},
\nnm\\
{D^\alpha}_{\beta(n)}&=\sum_{k=2}^{n}{n \choose k}{H^\alpha}_{\gamma(k)}{D^\gamma}_{\beta(n-k)},
\nnm\\
{C^\alpha}_{\beta(n)}&=-n\,{D^\alpha}_{\beta(n)}-\sum_{k=2}^{n-2}{n \choose k}{H^\alpha}_{\gamma(n-k)}{D^\gamma}_{\beta(k)}.
\end{align}

A recursion relation for our $E$ (their $\lambda$) is given by OW (4.11, 4.12), following from the transport equation OW (3.17).  We have found an alternate route to $E$, by first finding $K$ from a `transport equation' involving horizontal and vertical covariant derivatives,
$$
{K^\mu}_\alpha=\left(\xi^\beta\nabla_{*\beta}-\xi^\beta\nabla_{\beta*}+1\right){H^\mu}_\alpha,
$$
as detailed in our companion paper \cite{companion} (see also Footnote \ref{exp_foot}), and then inverting Eq.~(\ref{Kdef}).  Introducing the auxiliary quantities $dH$, the resultant recursion relations are
\begin{align}
{dH^\alpha}_{\beta(n)}&=-\frac{n-2}{n}\sum_{k=0}^{n-3}{n-3 \choose k}\Big[{\kappa^\alpha}_{\gamma(n-k)}{H^\gamma}_{\beta(k)}
\nnm\\&\qq\qq\qq\qq+{\kappa^\alpha}_{\gamma(n-k-1)}{dH^\gamma}_{\beta(k+1)}\Big],
\nnm\\
{K^\alpha}_{\beta(n)}&=(n+1){H^\alpha}_{\beta(n)}-n\;{dH^\alpha}_{\beta(n)}.
\nnm\\
{E^\alpha}_{\beta(n)}&=\sum_{k=0}^{n}{n \choose k}{D^\alpha}_{\gamma(n-k)}{K^\gamma}_{\beta(k)}.
\end{align}

\subsection{Parallel-propagator first derivatives\\ and `second-order Jacobi propagators'}

We can write the two first derivatives of the parallel propagator and the three bitensors of Eqs.~(\ref{LJI}), which one might call the `second-order Jacobi propagators,' as
\begin{align}
g_{\mu\alpha;\nu}&={g^\beta}_\mu{g^\gamma}_\nu A_{\alpha\beta\gamma},
\nnm\\
g_{\mu\alpha;\gamma}&={g^\beta}_\mu B_{\alpha\beta\gamma},
\nnm\\
{I^\mu}_{\beta\gamma}&={g^\mu}_\alpha {I^\alpha}_{\beta\gamma},
\nnm\\
{J^\mu}_{\beta\gamma}&={g^\mu}_\alpha {J^\alpha}_{\beta\gamma},
\nnm\\
{L^\mu}_{\beta\gamma}&={g^\mu}_\alpha {L^\alpha}_{\beta\gamma},
\end{align}
The bitensors $A$ and $B$ coincide with those of OW, and they do not discuss $I$, $J$, and $L$.

Defining the quantities
$$
{\mc R}_{\alpha\beta\gamma(n)}=R_{\alpha\beta\delta_1\gamma;\delta_2\ldots\delta_n}\xi^{\delta_1}\ldots\xi^{\delta_n},
$$
the recursion relations for $A$ and $B$ are
\begin{align}
A_{\alpha\beta\gamma(n)}&=-\frac{n}{n+1}\mc R_{\alpha\beta\gamma(n)}
\nnm\\
&\quad-\frac{1}{n+1}\sum_{k=0}^{n-2}{n \choose k}A_{\alpha\beta\delta(k)}{C^\delta}_{\gamma(n-k)},
\nnm\\
B_{\alpha\beta\gamma(n)}&=\frac{1}{n}\sum_{k=0}^{n}{n \choose k}A_{\alpha\beta\delta(k)}{D^\delta}_{\gamma(n-k)}.
\end{align}

Though recursion relations for $I$, $J$, and $L$ could be constructed via Eqs.~(\ref{LJI}a) using those for the world-function third derivatives (some of which are given by OW), or via Eqs.~(\ref{LJI}b), we have found it most efficient to use the relations 
\begin{align}
{I^\mu}_{\beta\gamma}&=\nabla_{*\gamma}{H^\mu}_\beta,
\nnm\\
{J^\mu}_{\beta\gamma}&=\nabla_{\beta*}{H^\mu}_\gamma,
\nnm\\
{L^\mu}_{\beta\gamma}&=\nabla_{\gamma*}{K^\mu}_\beta
\nnm\\
\end{align}
(see Footnotes \ref{exp_foot} and \ref{sojp_foot} and Ref.~\cite{companion}).  The resultant recursion relations, introducing several auxiliary quantities along the way, are
\begin{align}
{\delta\kappa^\alpha}_{\beta\gamma(n)}&={R^\alpha}_{(\gamma|\beta|\delta_1;\delta_2\ldots\delta_n)}\xi^{\delta_1}\ldots\xi^{\delta_n},
\\
{\Delta\kappa^\alpha}_{\beta\gamma(n)}&={R^\alpha}_{\delta_1\beta\delta_2;\delta_3\ldots\delta_n\gamma}\,\xi^{\delta_1}\ldots\xi^{\delta_n},
\nnm\\
\delta g_{\alpha\beta\gamma(n)}&=\sum_{k=0}^{n}{n\choose k}A_{\alpha\beta\delta(n-k)}{H^\delta}_{\gamma(k)},
\nnm\\
\Delta g_{\alpha\beta\gamma(n)}&=B_{\alpha\beta\gamma(n)}+\sum_{k=0}^{n}{n\choose k}A_{\alpha\beta\delta(n-k)}{K^\delta}_{\gamma(k)},\nnm
\end{align}

\begin{widetext}
\begin{align}
{\delta H^\alpha}_{\beta\gamma(n)}&=-\frac{n}{n-2}\sum_{k=0}^{n-1}{n-1\choose k}\left[\frac{n-k+1}{n+1}{\delta\kappa^\alpha}_{\delta\gamma(n-k)}{H^\delta}_{\beta(k)}+\frac{k}{n+1}{\kappa^\alpha}_{\delta(n-k+1)}{\delta H^\delta}_{\beta\gamma(k-1)}\right],
\nnm\\
{\Delta H^\alpha}_{\beta\gamma(n)}&=-\frac{n-1}{n+1}\sum_{k=0}^{n-2}{n-2\choose k}\left[{\Delta\kappa^\alpha}_{\delta\gamma(n-k)}{H^\delta}_{\beta(k)}+{\kappa^\alpha}_{\delta(n-k)}{\Delta H^\delta}_{\beta\gamma(k)}\right],
\nnm\\
{\Delta dH^\alpha}_{\beta\gamma(n)}&=-\frac{n-2}{n}\sum_{k=0}^{n-3}{n-3\choose k}\Big[{\Delta\kappa^\alpha}_{\delta\gamma(n-k)}{H^\delta}_{\beta(k)}+{\kappa^\alpha}_{\delta(n-k)}{\Delta H^\delta}_{\beta\gamma(k)}
\nnm\\
&\qq\qq\qq\qq\qq\quad+{\Delta\kappa^\alpha}_{\delta\gamma(n-k-1)}{dH^\delta}_{\beta(k+1)}+{\kappa^\alpha}_{\delta(n-k-1)}{\Delta dH^\delta}_{\beta\gamma(k+1)}\Big],\nnm
\end{align}
\begin{align}
{I^\alpha}_{\beta\gamma(n)}&={\delta H^\alpha}_{\beta\gamma(n)}+\sum_{k=0}^{n}{n\choose k}\delta g\indices{_\delta^\alpha_{\gamma(n-k)}}{H^\delta}_{\gamma(k)},
\nnm\\
{J^\alpha}_{\beta\gamma(n)}&={\Delta H^\alpha}_{\gamma\beta(n)}+\sum_{k=0}^{n}{n\choose k}\Delta g\indices{_\delta^\alpha_{\beta(n-k)}}{H^\delta}_{\gamma(k)},
\nnm\\
{L^\alpha}_{\beta\gamma(n)}&=(n+1){\Delta H^\alpha}_{\beta\gamma(n)}-n\;{\Delta dH^\alpha}_{\beta\gamma(n)}+\sum_{k=0}^{n}{n\choose k}\Delta g\indices{_\delta^\alpha_{\gamma(n-k)}}{K^\delta}_{\beta(k)}.
\end{align}

The explicit results needed to write out the $O(\xi,\dot\xi)^4$ GDE [from Eq.~(\ref{gde_exact})] are

\begin{align}\nnm
{K^\mu}_\beta=&\;{g^\mu}_\alpha\bigg[{\delta^\alpha}_\beta-\tfrac{1}{2}{R^\alpha}_{\xi\beta\xi}-\tfrac{1}{6}{R^\alpha}_{\xi\beta\xi;\xi}
+\tfrac{1}{24}\Big(-{R^\alpha}_{\xi\beta\xi;\xi\xi}+{R^\alpha}_{\xi\gamma\xi}{R^\gamma}_{\xi\beta\xi}\Big)
\\\nnm
&\qq\qq+\tfrac{1}{120}\Big(-{R^\alpha}_{\xi\beta\xi;\xi\xi\xi}+3{R^\alpha}_{\xi\gamma\xi;\xi}{R^\gamma}_{\xi\beta\xi}+{R^\alpha}_{\xi\gamma\xi}{R^\gamma}_{\xi\beta\xi;\xi}\Big)+O(\xi^6)\bigg],
\\
{H^\mu}_\beta=&\;{g^\mu}_\alpha\bigg[{\delta^\alpha}_\beta-\tfrac{1}{6}{R^\alpha}_{\xi\beta\xi}-\tfrac{1}{12}{R^\alpha}_{\xi\beta\xi;\xi}
+\tfrac{1}{120}\Big(-3{R^\alpha}_{\xi\beta\xi;\xi\xi}+{R^\alpha}_{\xi\gamma\xi}{R^\gamma}_{\xi\beta\xi}\Big)+O(\xi^5)\bigg],
\end{align}

\begin{align}\nnm
{L^\mu}_{\beta\gamma}=&\;{g^\mu}_\alpha\bigg[{R^\alpha}_{\beta\xi\gamma}+\tfrac{1}{2}\Big({R^\alpha}_{\beta\xi\gamma;\xi}-{R^\alpha}_{\xi\beta\xi;\gamma}\Big)
+\tfrac{1}{6}\Big({R^\alpha}_{\beta\xi\gamma;\xi\xi}-{R^\alpha}_{\xi\beta\xi;\xi\gamma}
-{R^\alpha}_{\beta\xi\delta}{R^\delta}_{\xi\gamma\xi}-3{R^\alpha}_{\delta\xi\gamma}{R^\delta}_{\xi\beta\xi}\Big)
\\\nnm
&\qq\qq+\tfrac{1}{24}\Big({R^\alpha}_{\beta\xi\gamma;\xi\xi\xi}-{R^\alpha}_{\xi\beta\xi;\xi\xi\gamma}+\{6{R^\alpha}_{\delta\xi\gamma;\xi}-{R^\alpha}_{\xi\xi\delta;\gamma}\}{R^\delta}_{\xi\beta\xi}
\\\nnm
&\qq\qq\qq\qq-3{R^\alpha}_{\beta\xi\delta;\xi}{R^\delta}_{\xi\gamma\xi}-{R^\alpha}_{\xi\xi\delta}{R^\delta}_{\xi\beta\xi;\gamma}
-{R^\alpha}_{\beta\xi\delta}{R^\delta}_{\xi\gamma\xi;\xi}+{R^\alpha}_{\delta\xi\gamma}{R^\delta}_{\xi\beta\xi;\xi}\Big)+O(\xi^5)\bigg],
\\\nnm
{J^\mu}_{\beta\gamma}=&\;{g^\mu}_\alpha\bigg[{R^\alpha}_{\gamma\xi\beta}+\tfrac{1}{6}\Big(3{R^\alpha}_{\gamma\xi\beta;\xi}-{R^\alpha}_{\xi\gamma\xi;\beta}\Big)
+\tfrac{1}{12}\Big(2{R^\alpha}_{\gamma\xi\beta;\xi\xi}-{R^\alpha}_{\xi\gamma\xi;\xi\beta}
+2{R^\alpha}_{\gamma\xi\delta}{R^\delta}_{\xi\beta\xi}+2{R^\alpha}_{\delta\xi\beta}{R^\delta}_{\xi\gamma\xi}\Big)
+O(\xi^4)\bigg],
\\
{I^\mu}_{\beta\gamma}=&\;{g^\mu}_\alpha\bigg[-\tfrac{2}{3}{R^\alpha}_{(\beta\gamma)\xi}+\tfrac{1}{12}\Big({R^\alpha}_{\xi\xi(\beta;\gamma)}-5{R^\alpha}_{(\beta\gamma)\xi;\xi}\Big)
+O(\xi^3)\bigg].\label{LJIexp4}
\end{align}

\end{widetext}

\bibliography{library}

\end{document}